# Progress in particle tracking and vertexing detectors

Nicolas Fourches (CEA/IRFU): 22nd March 2020, University Paris-Saclay

## Abstract:

This is part of a document, which is devoted to the developments of pixel detectors in the context of the International Linear Collider. From the early developments of the MIMOSAs to the proposed DotPix I recall some of the major progresses. The need for very precise vertex reconstruction is the reason for the Research and Development of new pixel detectors, first derived from the CMOS sensors and in further steps with new semiconductors structures. The problem of radiation effects was investigated and this is the case for the noise level with emphasis of the benefits of downscaling. Specific semiconductor processing and characterisation techniques are also described, with the perspective of a new pixel structure.

# TABLE OF CONTENTS:

a) Radiation hardness design rules in the layout of the pixel (this was made in most designs)

b) Radiation hard technologies such as SOI or more recently FDSOI

### 2.3.4. Simulation of pixel structures

## 2.4. The second step forward: other structures studied

1. First design a pixel with much improved spatial resolution by downscaling
2. Second : drastic radiation hardness improvements are sought
3. Use of device simulation is the preferred tool for this purpose

### 2.4.1. Particle transport through silicon: simulation

### 2.4.2. Noise in pixels: computation and simulation

a) W decrease: capacitance decrease transconductance increase , possible sampling rate increase., possible leakage current decrease

b) L decrease: capacitance decrease transconductance increase, possible sampling rate increase. possible leakage current decrease

c) Temperature decrease: Johnson an Flicker Noise decrease , leakage current decrease

### 2.4.3. Global pixel design

a) Pixel size or pitch , lateral dimensions

b) Number of hits per unit are and per seconds.

a) Data flow (assuming one bit pixels)

b) Number of pixels per identical array

c) Address length in bits

### 2.4.4. A new pixel concept

a) Ability to trap or to localize carriers

b) Ability to discriminate between carriers

a) The Detection Mode in which the Upper gate is negatively biased Vgate=Val <0 . In this mode, Vdd and Vss can be either grounded or kept at Vgate.In this mode the transistor is off

b) The Readout Mode: the upper gate is positively biased to switch the transistor in the On mode. The Drain is positively biased and the source is biased in current mode.This is the only mode where power is dissipated.

c) The reset mode in which the source is negatively biased and such is the case for the bulk and the drain grounded. An electron flow from the source is injected in the whole transistor and through the QW with holes recombining. The gate is positively biased. The substrate is grounded so that the electrons flow through the buried gate.

d) Another mode exists in which the Upper gate, Drain Source are grounded and so is the bulk, the holes in the QW should remain trapped and no increase due to ionizing particles should occur.

### 2.4.5. Simulation techniques and physics

a) Schrödinger-Poisson model which is a recursive method to determine the Eigenfunctions if the system is steady state this is accurate but is time consuming especially in 3D. It is not adapted to transient simulations for this reason, because it would need solving the coupled Schrodinger-Poisson equations at each step. This approach was one of the first used for quantum well modelling. This may be used in 3D at the expense of simulation time.

b) Density gradient model: this model is based on a transport equation derived from a quantum (effective) potential $\Lambda$.

$$\vec{J}_n = qD_n \nabla n - qn\mu_n \nabla(\psi - \Lambda) - \mu_n n (kT_L \nabla(\ln n_{ie}))$$

The so-called Wigner distribution function is used in this case defined as

### 2.4.6. Associated technologies

### 2.4.6.1. Associated technologies : ion implantation

a) 1MeV Zn ion implantation for peak concentrations of $10^{18}$ cm$^{-3}$.

b) 1 MeV Ge implantation for a peak concentration above $6 \times 10^{21}$ cm$^{-3}$

c) 14 MeV P implantation for a peak concentration of $10^{14}$ cm$^{-3}$.

Goal

Ion

Substrate

Energy

Doses

Substrate temperature

Implanted layer

Q well

Ge Implantation

High resistivity silicon wafer

100 ohm.cm

1MeV

Room temperature (300 K)

n-type buried layer

P implantation

High resistivity silicon wafer

100 ohm.cm

High energy

14 MeV

*Room temperature (300 K)*

Trapping layer

Implanted layer

Zn implantation

High resistivity silicon wafer

100 ohm.cm

25 µm @ 5V SCZ 400pF/cm$^2$

1 MeV

Room temperature (300 K)

Implanted layer

Q well

Ge Implantation

Low resistivity silicon wafer 1 ohm cm

0.8 µm @ 5V SCZ 13 nF /cm$^2$

1MeV

Room temperature (300 K)

n-type buried layer

P Implantation

Low resistivity silicon wafer

High energy 14 MeV

Room temperature (300 K)

Trapping layer

Implanted layer

Zn Implantation

Low resistivity silicon wafer

1 MeV

Room temperature (300 K)

a)  $3 \times 10^{17}$ cm$^{-2}$ with an introduction rate of approximately $2 \times 10^4$ cm$^{-1}$ this corresponds to a peak concentration is $6 \times 10^{21}$ cm$^{-3}$.

b)  The SIMS measured dose was $2.29 \times 10^{17}$ cm$^{-2}$ this corresponds to a peak concentration of $5.8 \times 10^{21}$ cm$^{-3}$. Or 12 % of the atomic density.

c)  We have measured such value below this target with the Oxygen incident ions and above this target with the Cs so that the peak concentration lies between these two values. The difficulties for implanting at such high fluxes and duration main explain the discrepancies.

d)  The positive MCs ions give higher value than the negative ones, MCs-

a)  The measured integrated flux was as expected $10^{14}$ cm$^{-2}$, using the electrode current. (see table )

b)  There is a discrepancy between the integrated flus measured with the Cs, which higher than that measured with oxygen ion.

c)  The peak concentration is on average: $1.83 \times 10^{18}$ cm$^{-3,}$ for Oxygen incident ions and $6.95 \times 10^{13}$ cm$^{-2}$ integrated flux, below the value both measured and targeted this is obtained on low resistivity samples. On the high-resistivity samples the peak concentration is: $1.2 \times 10^{18}$ cm$^{-3}$ corresponding to : $5.45 \times 10^{13}$ cm$^{-2}$

d)  For Cs incident ions, the results on HR samples show an average integrated flux of: $1.22 \times 10^{14}$ cm$^{-2}$ and a peak concentration of: $2.15 \times 10^{18}$ cm$^{-3}$, above the measured values by 20 %. This could be explained by the presence in the secondary ions of species with the same Charge to Mass Ratio of the Ge secondary ions. Otherwise, although lower, the peak concentration corresponds to the target value by -11% to -24 % that is very good, if we consider that the implantation is inhomogeneous, which is probed by the electrode current.

a)  The average measured integrated flux with Oxygen SIMS was: $3.91 \times 10^{12}$ cm$^{-2}$

b)  The corresponding peak concentration was measured on average: $6 \times 10^{16}$ cm$^{-3}$ that is comparable to the target value obtained using the target implantation and the introduction rate derived from SRIM simulations. No measurements with Cs as the incident ions were made here.

**2.4.6.2.    Associated technologies : characterization**

a)  Raman spectroscopy: this technique is non-destructive and give information on the coupling of the phonons with incident light.

b)  Deep Level Transient Spectroscopy, which involves a MOS, PN, PIN, Schottky structure and give quantities related to the electrical properties of defects, capture cross-section, energy levels and densities.

a)  Ge-Si mode at 401 cm$^{-1}$ : $\gamma=1.2$ +/-0.1

b)  Ge-Ge mode at 297 cm$^{-1}$ : $\gamma=1.13$ +/0.1

c)  Si-Si mode at 520 cm$^{-1}$ : $\gamma=1.13$ +/-0.11 in silicon.

a)  The shift of the 520 cm$^{-1}$ Si-Si line due to the strain in the upper part of the region (tensile in this case , shift to lower wavevectors) and the disorder.

b)  The Ge-Si line which is due to the buried layer with high germanium concentration which is compressively strained

### 2.4.7. Future: Epitaxial growth and outlook
### 3. Conclusions:

_________________________________________________________________

In almost all accelerator based particle physics experiments, the reconstruction of charged particle tracks and vertices is necessary [1] [2]. The magnetic field in the inner detector enables the determination of the particles impulsions and therefore access to a physical quantity, needed to characterize the decay process of the particles generated in the collision. It is also necessary to determine with great accuracy the corresponding vertices. The term vertex should be understood as follows. This is the point where the trajectories have their origin. Figure 1 makes a short description of how to make the needed reconstruction. We can also define the Impact Parameter as being the distance between the IP (Interaction Point) and the start of the trajectory of the charged particle. We will show that the pixel detector we design can improve the resolution for measuring this quantity see ILC reference design report, Detectors, 2007 [3] [4] [5]. The Impact Parameter is the shortest distance between the primary vertex and the track of the charged particle as it passes close to it.

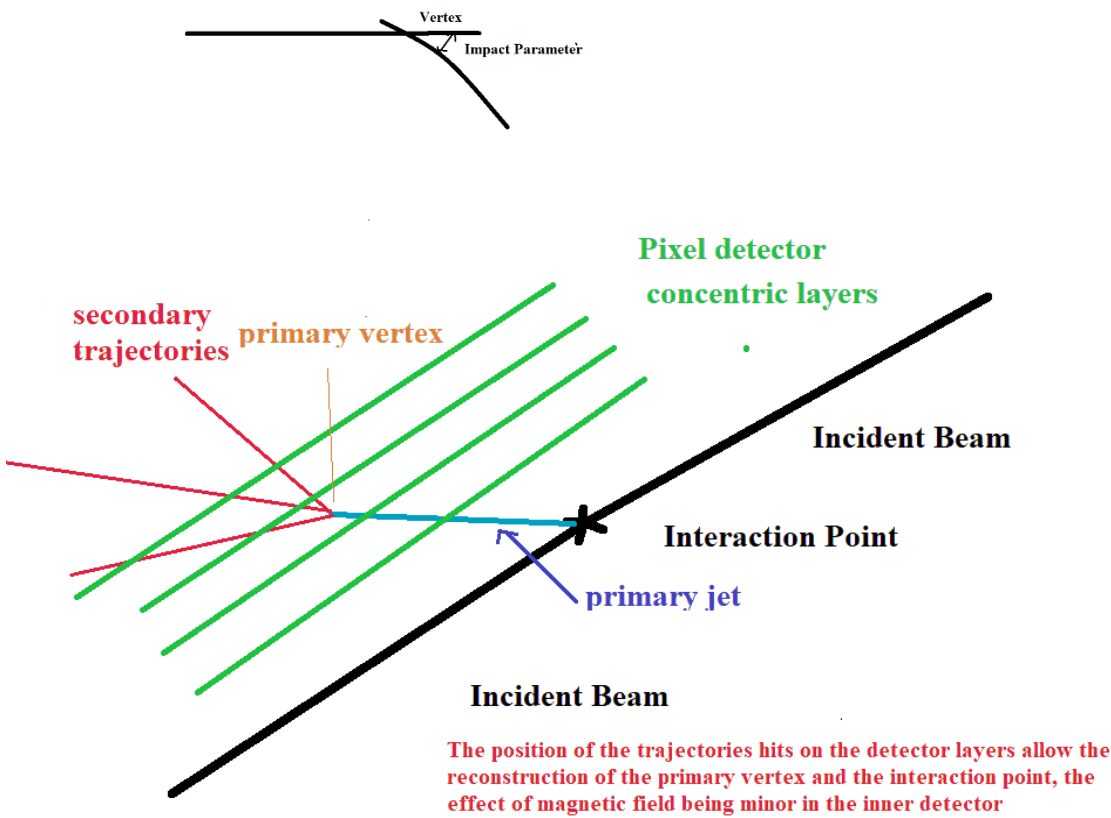

Fig. 1: the simplified description of how a vertex detector operates. The incident beams make an angle of alpha between each other to maintain an interaction volume set to the needs.

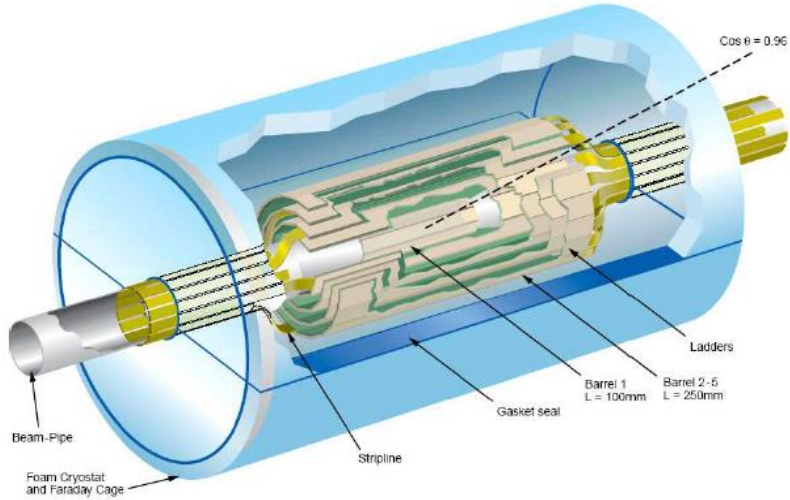

FIGURE 5.1. 'Long-barrel' option for the ILC vertex detector. The cryostat is an almost massless foam construction, and has a negligible impact on physics performance.

Fig. 2: Sketch of the pixel detector with the concentric layers: the thickness of the layers should be kept below a few hundreds of microns (ILC Reference Design Report [4] [5] [6]).

In many cases, we need to determine the position of the interaction and to tag the bunch in which the interaction has occurred. This is possible with another detector, which is then time sensitive. The bunch can be tagged when getting out of the pipe with a transformer (it is made with charged particles of the same nature [7]) . Hence, with this in mind the readout electronics is should be fast with timing resolution well below the ps, it should need a fast outer tracker to determine the stop signal with a resolution attaining these values.

With a Position Sensitive Detector, the constraints are different. The position of the Interaction Point can be estimated, using the secondary vertex and the estimation of primary trajectory, giving the primary vertex. In the case of the so-called heavy flavour physics, the primary vertex is close to the interaction point, and the reconstructed tracks can be used to estimate an Impact Parameter. In our case, with very high point to point resolution the Impact Parameter can be estimated with a better resolution than in (ILC reference design report IV-117 [7])

This may be made using the energy and the momenta of the secondary particles when the primary particle is neutral or better if the primary particle is charged we can use its track reconstruction with the utmost inner layers of the pixel detector. The determination of the Interaction Point using this technique when possible should be more accurate than proposed timing methods [8] [9]. Let us say the velocity of the particles in the beam, is $c/10$, then $v=3x10^9$ cm.s$^{-1}$. Then if we can achieve a spatial resolution of $1\mu m=10^{-4}$ cm. The corresponding time spread is $1/3x10^{-4-9}$ seconds, so the accurately: $3.3x10^{-14}$ = 33 fseconds. A timing resolution well below the picosecond would be necessary [10] [11]. This is a challenge for timing based detectors [12] especially because time tagging is necessary and not simply timing, today the state of the art time resolution is well above 1 ns in current PET for example, also 50 ps range Si detectors have been tested, and sampler chip down to this value exist. For this reason, the need of a good pixel-state of the art detector is a prerequisite. Up to now, resolutions close to the micrometre were achieved with monolithically integrated pixel detectors [13]. We will review most of the key developments until now and make the case for monolithic pixel defectors.

## 1. The trend towards 1-micron point-to-point resolution and below

The reconstruction of tracks (or trajectories) is prerequisite for particle tagging and identification. The detectors used for this purpose are either solid state or gaseous, liquid detectors are not used for tracking purposes because of the too high material budget they represent for trackers (ref high granularity calorimeter put ref with cold electronics)  We will review shortly the operating principle of first gaseous and second solid-state detectors.

### 1.1. Gaseous detectors :

In recent times, a lot of activity has turned back on the gaseous detectors for tracking purposes. The principle used here is the Time Projection Chamber Method. The tracks ionize the gas (with the appropriate mixture), ions and electrons are created and with the help of an applied electric field are separated and drift through the chamber towards the electrodes. Consequently, a displacement current appears on the electrodes (this can be computed using the well-known Shockley-Ramo theorem). The signal can then be processed electronically, and digitalized.  The advantage of such detectors is the apparent simple design and fabrication. As it is usually the case for such detectors, an internal amplification is implemented. The use of a grid and an electrodes bias at a critical potential induces an avalanche with the multiplications of carriers. This is the principle of GEMs (Gas Electron Multiplier [14]).

The MICROMEGAS MicroMesh [15] Gaseous type detectors were developed at IRFU for some experiments (ref needed). However, due to the superposition of tracks and their consequence (high number of ionized   species within the detecting volume), these chambers are not effective at high counting rates as the drift time of the charges is large. Space charge effects are very important [16]. Taking into consideration that the charges have to drift along a distance longer that a meter if we consider for instance the TPC proposed for the ILC tracker, this is a slow detector compared  to silicon trackers.  However, the performances of these detectors in terms of spatial resolution have improved, due to the microfabrication techniques that have been introduced. We will not comment more on these detectors as they have been investigated recently in many studies for both their applications and their operation principle.

### 1.2. Liquid-state based  detectors :

The liquid based detectors are not very practical for tracking purposes as their material budget is close to that of a solid and they have the disadvantages of gaseous detectors. High granularity liquid argon calorimeters [17] have some tracking possibilities. (DMILL developments on cold electronics [18] [19] [20] [21] [22] [23].

### 1.3. Solid-state based detectors :

As we have stated in the introduction the solid-state detectors are the most numerous solution to tracking and vertexing. Apart from gamma ray tracking where germanium is used as a detecting media in a bulk style device [24] [25] [26] most are made of thin layers of semiconductors and the detection of charged particles is made through a LET process [1]. Track reconstruction is made using a pixelated structure. The detector structure is practically always made of a concentric cylinders of pixel-detectors with the cylinders having increasing radii. One of the drawbacks of such designs is that they are exposed to high fluxes of charged or neutral particles that induce crystallographic defects in the materials constituting the detectors or charge the oxides used in silicon technologies [27]. I have recently developed a R&D on this particular subject as  the use of monolithic pixels is gaining importance recently, partly because they allow the optimum point-to-point resolution, down to the micron range that are not at the reach of other designs. This pixel family are the purpose of the next paragraphs.

## 2. The solution: the monolithically integrated pixel detector:
### 2.1.     Advantages and drawbacks

Eric Fossum introduced the principle of such detector for visible light imaging although researchers introduced earlier the terminology (Active Pixel Sensor or CMOS sensors [28] [29]). Active Pixel Sensors were introduced as charged particle of X ray photon detectors in 2001 [30] [31]. The operation principle of the early CMOS sensors is based on a 3T scheme with a photodiode operating in a quasi-photovoltaic mode. The photo-generated carriers diffuse through the PN junction made of an n electrode made with the diffused area of a P-MOS structure, which is n-type and the p type substrate, which is grounded. The n-type electrode is connected to the upper gate of an n-channel transistor and to the source of an n-channel transistor that acts as a reset switch. The amplifying transistor operates as a source follower. More elaborated designs have been proposed and tested but the basic principle has remained identical MIMOSA8) [32] [33] [34] [35] [36] [37] [38]. One of the drawback of the diffusion-mode charge-collection is its sensitivity to radiation-induced defects, particularly to point and extended defects located in the bulk silicon. Specific studies I have made on this subject show that carrier transport through diffusion is more sensitive than carrier transport through drift, even though the carrier collection length is larger [39]. To mitigate these effects, MAPS pixels implemented on high resistivity substrates were implemented and fabricated resulting in a depleted sensitive volume. As forecast, their radiation hardness (to massive non-ionizing particles inducing NIEL that contribute to atomic displacements) is better than the MAPS predecessors [1] [40]. The other effect of diffusion is to increase multiplicity (number of pixel sets per impinging particle on one pixel [41]).

The problems encountered in present experiments is the presence of multiple hits on single pixels. This will preclude the use of the today-designs, which are either too large in terms of spatial dimensions or too slow to tag efficiently the particle bunch that corresponds to the hit event. Therefore, the only way to progress is to either obtain very small pixels down to the squared micron scale, or/and have a time tagging pixel. These two approaches are complementary, although timing pixels are more difficult to design. The other issue is the fact that timing pixels have application in medical physics, making this area of research more attractive.

Important conceptual progress was made with the proposal of Quantum-Well Ge based pixels that reduce to a single n-channel transistor with a buried QW gate [42]. This is still at the conceptual and simulation level but technological progress is under way to obtain a viable process at the silicon level.

## 2.2. Spatial resolution : experimental physics requirements
### 2.2.1. Detection Physics at colliders
#### 2.2.1.1. Track reconstruction

Up to now, the tracker and vertex detectors are made of silicon strip or pixels sensors with a hybrid design resulting in a pin structure connected to Read Out Chip. The dimensions of these devices often exceeds the 50 µm x 50 µm squared. The technology used is based on a bump bonding procedure. The physics needs at high repetition rates has imposed such design at the LHC for instance. First, the outer trackers are necessary to determine the value of a charged particle momentum by reconstructing its trajectory from the position of the hits on the tracker layers[1]. The magnetic field inside the detector induces a curved track due to the Lorentz force to put it simply. In special relativity, the following classic field tensor can be used (no spin considered):

$$F^{\mu\nu} = \begin{pmatrix} 0 & -E_x/c & -E_y/c & -E_z/c \\ E_x/c & 0 & -B_z & B_y \\ E_y/c & B_z & 0 & -B_x \\ E_z/c & -B_y & B_x & 0 \end{pmatrix}$$

This tensor is valid when the electric field is nil Ex=Ey=Ez=0. This enables an accurate determination of the transverse momentum and if the detector has forward layers, it allows the characterization of events with low transverse momentum (θ low or high η=pseudorapidity Note 2). Hence, this procedure allows the determination of transverse energies, which are crucial for the analysis of events, especially where missing energies are important. These quantities are necessary for charged particle identification and mass determination.

## 2.2.1.2. Constraints on detector design

Most of the problems in track reconstruction arise from the number of tracks in the detecting media generated by the collisions is very important at high luminosity (number of primary collisions per cm$^2$ per second). This is particularly serious in modern hadronic experiments such as the LHC. The consequences are as follows:

a) Multiple interaction points in the incident colliding particle bunches
b) Multiple hits in single pixels even in the outer layers
c) Large NIEL (Non Ionizing Energy Loss) in the pixels leading to displacement defects in the silicon layers
d) Cumulative ionization in the solid state detectors leading to a total dose above 1 MGy in the operating time of the machine

The former constraints have a direct effect on data analysing and the latter on detector design. Let us analyse the ways these problems can be mitigated.

a) First reducing the bunch length and beam diameter would significantly limit the number of spurious interaction points. This is the trend, which is an objective at future ILC experiments [43] [3].
b) Increasing the granularity of the pixels layers inducing a reduction of multiple hits (per unit time) in each pixel ( this is the goal of this R&D)
c) Time-tagging in the outer layer pixels could help in reconstructing the appropriate tracks belonging to the appropriate interaction point. This is particularly difficult in the ps range due to the pixel detectors not the on-chip electronic.
d) Having the possibility to resolve the tracks using very small pixels may be an alternative because it reduces the ambiguities and thus enables the separation of tracks initiating from different interaction points.

One point should be explained. The use of small pixels with a thin detecting region drastically reduces the charge spread in the neighbouring pixels. The charge spread involving multiplicities is often used (and we have used it see Y. Li [41]) to make Centre Of Gravity determinations. These COG methods can improve the precision and resolution of the determination of the particle hit. However, this method requires that a pixel cluster should be considered, with no multiple hits (from other particles). This means that it is inefficient when close tracks are considered and where there is an occurrence of multiple hits. The only way to improve point-to point resolution is to use small pixels with no charge spread. Reduction of charge spread is sought by full depletion of the active/ detecting layer and reduction of its thickness down to a few microns, with drift being the dominant transport mechanism [39]. In this case the aspect ratio of the structure limits the charge spread, especially if the pixel is separated from the others by trenches or are made up in mesa like structure. The only case where two or more pixels can be hit by the \same particle is when the tracks is inclined. This can be mitigated by tilting the pixel-arrays in the forward and backward position in the inner detector (tilted pixels.).

These constraints are valid for the inner detector (tracker) in the outer and inner layers. However, for the vertex detector, which is the closest to the beam and interaction point other criteria are necessary.

## 2.2.2. The case for a micro-vertex detector

The physics after the primary collision requires when this is possible to identify the particles and this is true for:

a) Neutral particles such as the Z and W (W can be charged) uncharged electroweak massive gauge bosons and also the H (Higgs) boson
b) Tag charged particles such as B and Taus. These are short-lived particles
c) Make precise measurement of missing transverse energy

d) Make precise measurements of all decay channel of the Higgs boson, particularly the one that has the most important branching ratio the decay into b b̄ , b b bar. O (80 %). As these topics are studied at the LHC, the more difficult would be to study the coupling of the Higgs with other particles.

The Yukawa coupling of the Higgs doublet to quarks and leptons give their masses and mixings values. For this purpose, the identification of neutral or short-lived particles is only possible by tagging the final states, or more intermediate states. Direct detection of W bosons (with a mass of 90 GeV) is not possible because they decay in very short time amount.

For these gauge bosons, the mean lifetime is lower that $10^{-25}$ seconds, so there very little hope to detect them directly with charged particle detector. The mean distance is (with light velocity) equivalent to: $3\times10^{-17}$ meters so approximately 0.03 femtometers, this means that direct detection is almost difficult.

For example, the decay of W+ is into hadrons so tagging them is one of the best way to proceed.

This lifetime amount is directly related to the width of the resonance. This can be seen an analogy with a resonant circuit with a pole. It is indeed. The decay of the Higgs into many charged particles with a short lifetime means that precise tagging is needed to disentangle the decay modes of the Higgs boson, for instance.

However for the tau particles having a higher lifetime $10^{-13}$ seconds they can have an average track as long as $3\times10^{-13+8} = 3\times10^{-5}$ m = 30 µm which could be measured with a precise VTX detector.

The ILC vertex detector would be necessary to study the different origins of the Higgs resonance [1] .

The search for extra dimensions [44] [45] [46] [47] [48] [49] [50] [51] [52], which is no standard model physics or beyond standard model is a topic which in some cases is related to the graviton question. If one of the purpose if to check if warped dimensions exists this would certainly lead to a vertex displaced or not well defined. The common view to detect the presence of extra dimensions using a missing transverse energy as some particle would be not be detected in the presence of such dimensions, at either the electroweak and other scales (see ILC reference design report, physics [43] ). Up to now 2018 the size (the radii) associated to extra dimensions has been set to a limit of a few tens of microns [1], from cosmological constraints

In the Randall-Sundrum theory the Kaluza-Kein dimension are warped, this means that some excitation would occur for some masses of the order of:

$m_n \simeq (n + \alpha/2 - 1/4)\pi k e^{(-\pi k R)}$ , with $ke^{(-k\pi R)} \sim$ TeV. where α is defined as

R is the spatial extension of the extra dimensions.

f0 ~ exp[(1/2 − cf )ky], where cfk is their 5D mass .

One of the paradigm of the ED problem is to detect the KK excitations, the SM fields can propagate through these extra dimensions if they are flat for instance. In some other models, the Universal Extra Dimensions that assume a specific SM field propagation. Hence, the lowest KK excitation is a Dark Matter candidate and is stable. . Some ATLAS [44] and CMS analysis have set a lower limit to 2.5 TeV. using top quark tagging technique. One should be aware that the target of the ILC or CLIC project is to obtain a beams radius of a few micrometer or less in order to limit the extension of the IP to a point-like domain. Another field of research that could benefit from a micro-vertex is the test of SUSY (Supersymmetry), through for example-displaced vertices.

### 2.2.3. Momentum and VTX determination :

We can now determine the accuracy of the momentum determination using a micrometre-range vertex detector, together with vertex determination. Let's take a vertex detector is made of 3 layers. For the VT detector planned for the ILC the inner layer is set at 1.0 cm from the beam. Let's get a layer at 1 cm ,2 cm and

3cm from the beam, with have three layers separated by one centimetre. We have made a simple simulation code. We make the hypothesis that the pixels hit distribution is normal and that we can fit the measured data in order to get the position of the VTX. We another set of two points and we get the other track. The results are in the following figures (ref: reconstruction simulation code I have developed in Scilab). See (Fig.3 and Fig.4). Usually, methods that are more elaborate are used for track reconstruction [53].

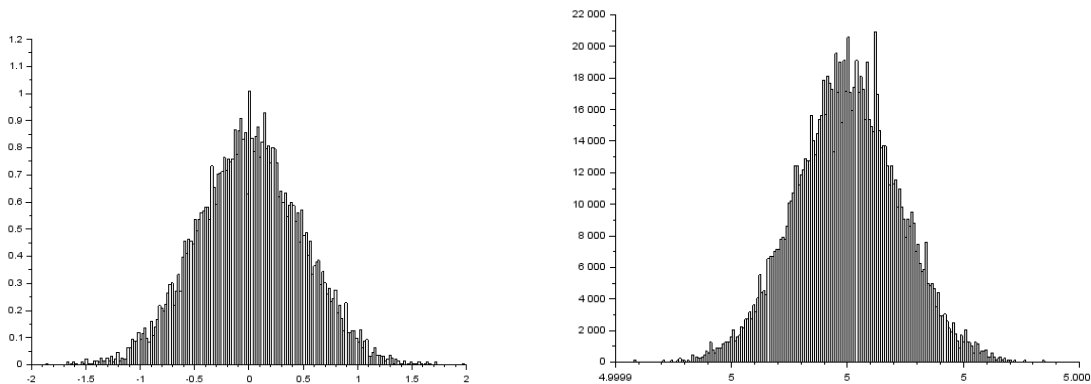

Fig. 3: Gaussian hit distribution: sigma=1 micron. Slope and abscissa at the origin (VTX). 10000 events.. The left distribution gives the abscissa at the origin and the right one the slope

The conclusion is straightforward the resolution at the VTX is one micron with either a uniform (tested here) or normal distribution. The resolution is directly related to the point-to-point resolution of the pixel array. We can improve this resolution by increasing the number of layers.

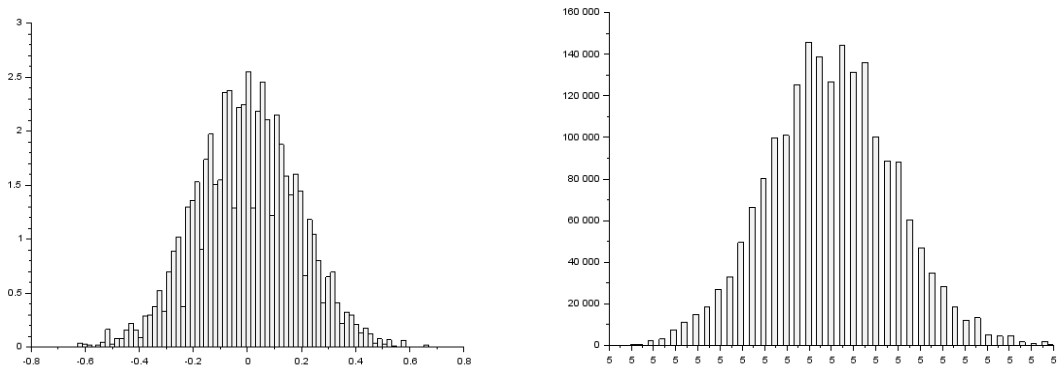

Fig.4: improved fitting Gaussian distribution 20000 events. The left distribution gives the abscissa at the origin and the right one the slope

In this case, the distribution is digitized as it could be by a binary pixel. From this simulation the single track resolution resolution defined as the FWHM is or the order of 0.4 micrometres. This shows that the deep submicron resolutions can be achieved with these methods and pixels design.

## 2.3. The first step forward: CMOS sensors

The initial idea to use of CMOS sensors in charged particle detection comes from the Strasbourg Group. We have joined our effort to get some pixel-arrays designed and fabricated in order to test this technology. This gave the MIMOSA series in which I was personally involved. We have studied the pixel readout and some MIMOSA architecture (MIMOSA 6-7-8-8bis [54] [31]. The work on the MIMOSAs were made with the contribution of the group at IRFU here with Pierre Lutz leading, and my coworkers, Yavuz Degerli that I chose and welcome first at a post-doctoral position and after at a permanent member of IRFU staff;. We note the contribution of Marc Besancon and F. Orsini. I led all the contributions to radiation effects and I had a major role in the choice of technologies and to the guidance of our PhD student (Yan Li [41] [55]).

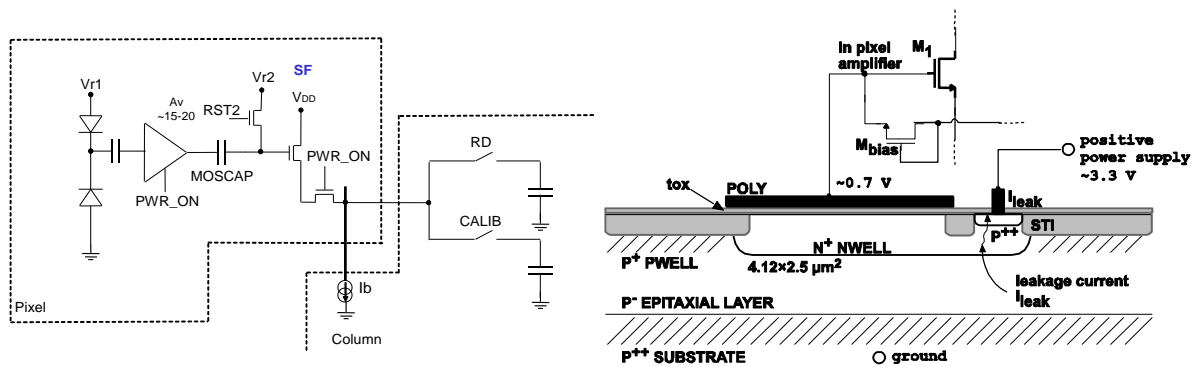

Fig.5. Pixel configurations and schematics of the MIMOSA series (MIMOSA 6). The sensitive part is the photodiode [54] [32] [31].

In the pixel the sensitive device if simply, a photodiode reversely biased at very low voltages. This means that the operation mode of the pixel is closer to a photovoltaic mode than a drift detector mode. This has important consequences in the properties of this kind of pixel. The presence of a non-depleted detecting zone lead to a slow collection of the generated charges, which are affected by the diffusion length in the material. We will study this in the following. The pixel is followed by a voltage amplifier, coupled to the sensitive node of the photodiode with a capacitor in most cases. The process used for implementation of this kind of pixel is a standard CMOS process at the technological node at the time of the study. A AMS 350 nm was first used and for the MIMOSA8 I proposed the 250 nm TSMC process for the pixel array we designed. This was a digital process, which made the implementation of the digital part of the array easier to implement [31]. Prior to that, we studied the possibility of digital pixels with a latched comparator. For the MIMOSA 8 series despite my recommendation the comparator had no current limitation, a current controlled latched has the advantage of being less noisy and offers the possibility of being insensitive to destructive latch up. The use of current controlled latches has been made on MIMOSA 16. The latched comparator was offset compensated so that this lead to a reduction of the Fixed Pattern Noise at least on un-irradiated chips.

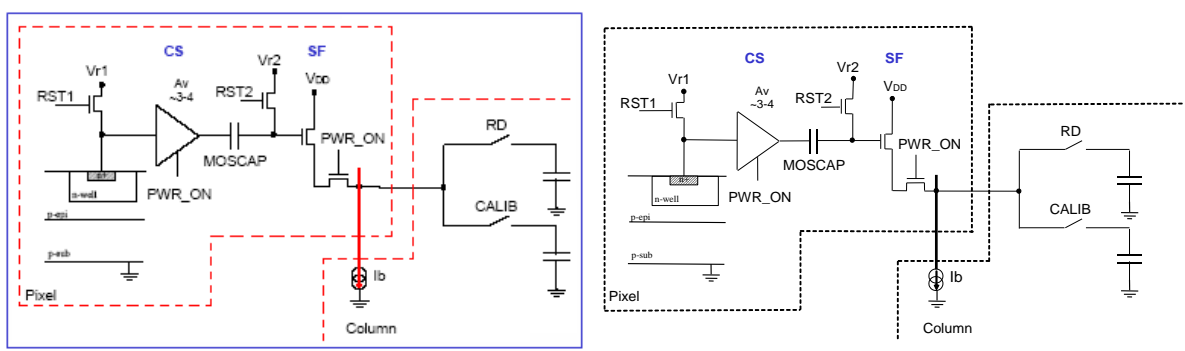

Fig.6: Pixel schematic implemented in the MIMOSA 8 pixel. The principle is still based on a photodiode. The output switch capacitors are present to reduce the FPN [31].

The development of a successive approximation ADC was a follow up of this R&D, in the following pixel-chips.

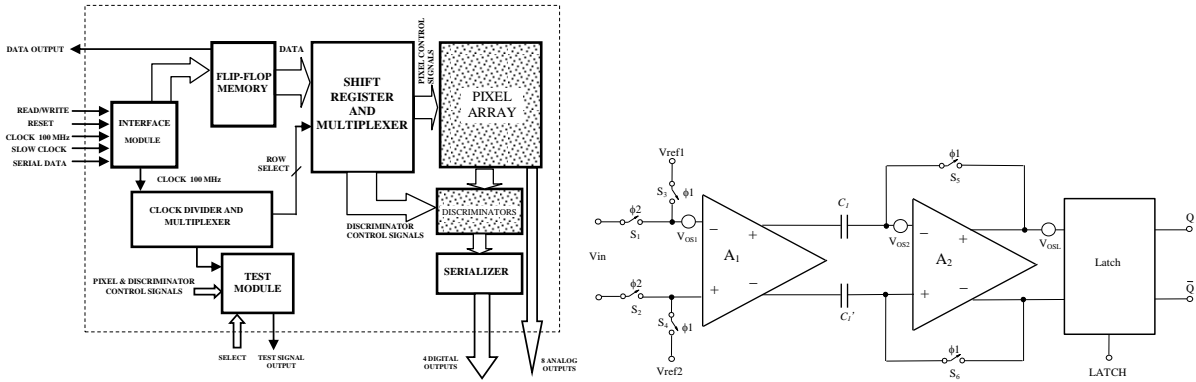

Fig.7: Architecture of the comparator system (offset compensated) for binary outputs and the digital part of the MIMOSA 8 chip [35].

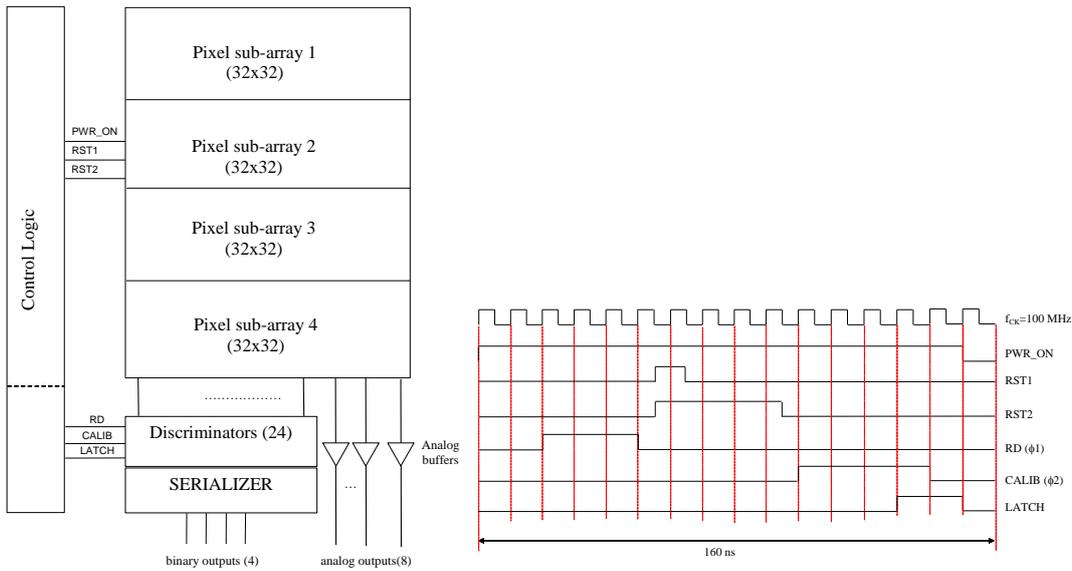

Fig.8: Architecture of a pixel array, with column circuit for data acquisition.

In order to characterize the properties of the pixels the arrays have included different pixel sizes and two readout modes: analogue and digital (one bit) using a discriminator. The description of the array can be found in Fig.8, with the corresponding chronogram for chip control. The signals from the digital pixel can be serialized and output to the external interface but the data is not compressed. The all chip can be programmed but this option has not been used in many of the tests described here. Programming only alters the duration of the different control signals. For the test of the chip two PC boards have been developed with a upper board designed with a hole necessary to irradiate the chip bonded directly on it.

### 2.3.1. Experimental results: X rays

Tests on MIMOSA6 have been disappointing. On the other hand, tests on the MIMOSA8 chip have been very interesting. The first result is that of the X rays on all the analog pixels. $^{55}$Fe X rays have a line at 5.9 keV and 6.4 keV. The results shown in Fig. 9 show that this kind of detector have a good spectro-imaging capability.

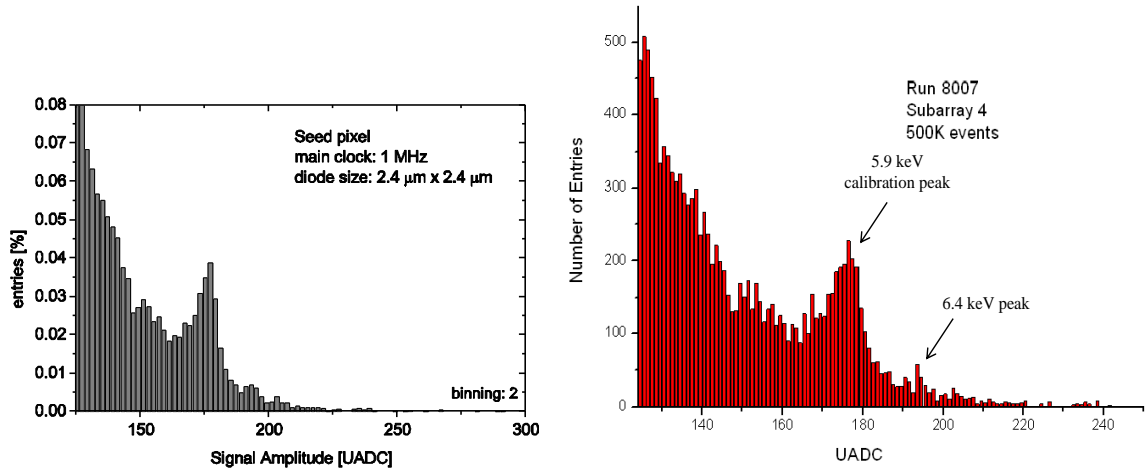

Fig.9: Histogram of hits with the X rays from a $^{55}$Fe source, both with clusterization and seed pixel. One can distinguish the two lines of the source 5.9 keV and 6.4 keV demonstrating the possible spectro-imaging potentiality [32] [31].

### 2.3.2. Experimental results: Beam tests

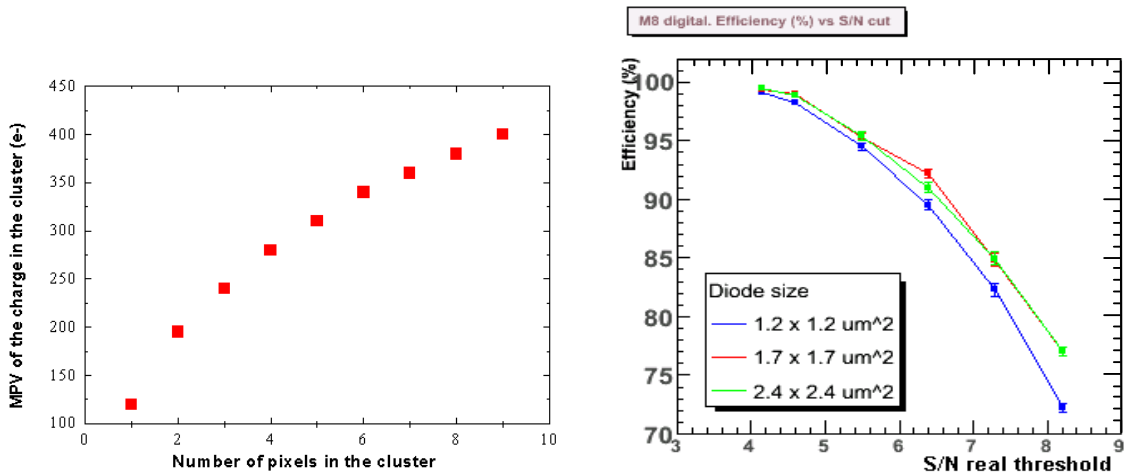

Fig.10: Beam test results with either e- from the DESY facility and p+ from CERN facility (130 GeV). Detection efficiency versus threshold and signal in the seed pixel.

Further experiments include the beam tests made with a telescope that can be used to evaluate the resolution of the pixel array. The description was made in [35] [32]. The main conclusions are as follow. The detection efficiency remains above 95 % for a signal to noise ratio set lower than five. This can be made by lowering the discriminator threshold. For the digital or the analysis on the analogue pixels. The second conclusion stems from the fact that the hit multiplicity defined by the number of pixels that flip for one particle hit is very high. This means that the charge spread is important and that a strong diffusion or random walk of carriers occurs. The estimation of the time necessary for charge collection is around 100 ns.

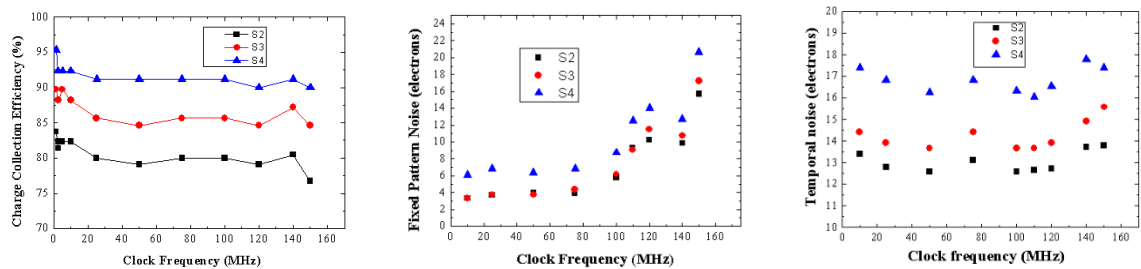

Fig.11: Charge collection efficiency versus clocking frequency. Contribution of temporal noise and fixed pattern noise [32] [31].

The Fig.11 shows that the arrays is still functional at 150 MHz clocking frequency that corresponds to approximately 10 MHz scanning frequency. This means that the corresponding time is 100 ns. The transport of charges limits the operation frequency not the readout.

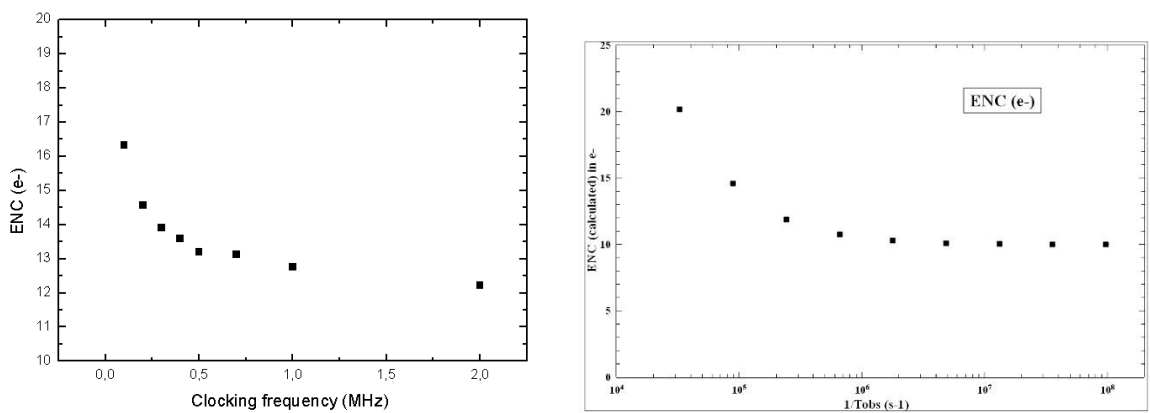

Fig.12: Calculation of the temporal noise for a 3T pixel, as a function of clocking frequency (see formulae) and experimental measured temporal noise for the MIMOSA pixel with comparable architecture, the noise is in Equivalent Noise Charge an corresponds to a temporal noise [56] [57].

I have investigated the variation of the noise expressed in Equivalent Noise Charge with respect to a so-called observation time (see paper for further details). This was computed analytically using the Campbell Theorem. I find a similar frequency dependence with the sampling frequency of the pixel despite having an architecture slightly different. The frequency here is the sampling frequency, which is the same as the clocking frequency of each pixel (and not the original clock). In this experiment, I have shown that classical calculation can be made which are still valid at low noise level of a few electrons.

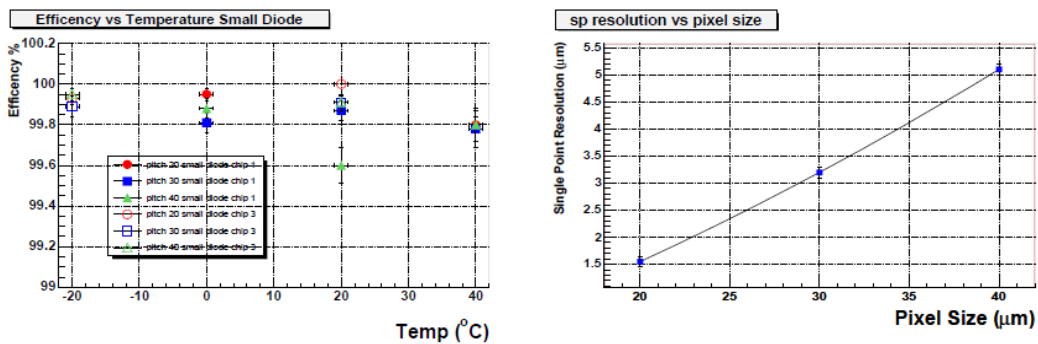

Fig. 13: detection efficiency versus temperature and spatial resolution versus pixel size for the analog outputs of a MIMOSA pixel array (MIMOSAs [58] [59]).

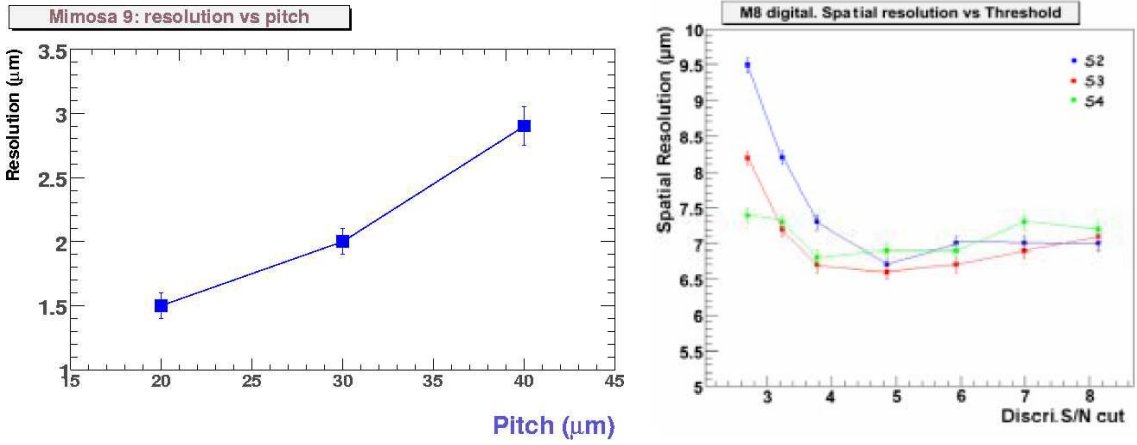

Fig.14: Spatial resolution versus pixel pitch with respect for analog pixels and digital MIMOSA 9 and 8 array as measured on the digital pixels (left and right) [60] [41]

Table 1: beam tests for the MIMOSA 8 pixel array [41]

| Beam Test | Sub Array | Temporal Noise (e-) | Charge in 3×3 cluster (e-) | S/N of seed pixel | Detection efficiency to MIPs |
|---|---|---|---|---|---|
| DESY 2005 | S2 (1.2 x 1.2 µm²) | 11±2 | 353±4 | 8.6±0.2 | 98.2±0.4 % |
|  | S3 (1.7 x 1.7 µm²) | 12±2 | 392±4 | 9.8±0.1 | 98.6±0.3 % |
|  | S4 (2.4 x 2.4 µm²) | 14±2 | 433±5 | 9.4±0.2 | 98.0±0.4 % |
| CERN 2006 | S2 (1.2 x 1.2 µm²) | 10±2 | 306±4 | 6.2±0.5 | 97.6±0.6 % |
|  | S3 (1.7 x 1.7 µm²) | 11±2 | 328±4 | 6.3±0.2 | 98.1±0.5 % |
|  | S4 (2.4 x 2.4 µm²) | 12±2 | 381±5 | 6.6±0.6 | 97.5±0.5 % |

Table 6.2: The beam test performance of the analog outputs for MIMOSA 8. Measured at DESY in 2005 and CERN at 2006.

The two above figures that correspond to a distinct work show that with the cluster method good resolution can be obtained at lest in the analog mode allowing a better Centre Of Gravity determination. With a pitch of 20 µm the resolution of the array is < 2µm. However this means that the charge spreads on at least 50 µm from the impact point . This is a huge value that poses a number of questions. It will induce unwanted pile up because of the size of the clusters necessary to reconsttruct the hit point, and be a problem at high luminosity. To circumvent that the answer is not simple. We have :

   a) To reduce the pixel size
   b) To limit the charge spread on neighbouring pixels

The first item is the trend that is followed from the beginning. But the question that stems from that is how to reduce lateral diffusion.

The simple solution are as follow.

   a) Limit the pixel aspect ratio , by limiting the sensitive thickness
   b) Use drift transport that can be oriented from the bottom to the top using the appropriate material and electrodes

One can find that the use of drift detector is a step back as it imposes large voltage bias when use on a significant thickness. However, many institutes have used HV CMOS to satisfy these needs. We can summarize the results from mimosa8 and his follower MIMOSA 16 below.

Table 2: comparison between the MIMOSAs [41] [61] [62] [63] [64]

| Chip | Sub array | CVF (µV/e-) | Input Referred Noise (ENC) | Input Referred FPN (ENC) | S/N ($^{55}$Fe) | Charge Collection Efficiency | |
|---|---|---|---|---|---|---|---|
| | | | | | | 3x3 | 5x5 |
| MIMOSA 8 | S2 (1.2 x 1.2 µm²) | 66 | 11 e- | 4 e- | 130 | 66% | 80% |
| | S3 (1.7 x 1.7 µm²) | 60 | 12 e- | 5 e- | 110 | 73% | 85% |
| | S4 (2.4 x 2.4 µm²) | 52 | 14 e- | 6 e- | 90 | 82% | 92% |
| MIMOSA 16 14 µm version | S2 (2.4 x 2.4 µm²) | 59 | 11 e- | 2 e- | 152 | 36% | - |
| | S3 (2.4 x 2.4 µm²) Rad. Tol. | 54 | 12 e- | 2 e- | 127 | 37% | - |
| MIMOSA 16 20 µm version | S2 (2.4 x 2.4 µm²) | 60 | 11 e- | 2 e- | 151 | 25% | - |
| | S3 (2.4 x 2.4 µm²) Rad. Tol. | 53 | 13 e- | 2 e- | 132 | 29% | - |

Table 6.1: Characteristics of MIMOSA 8 and MIMOSA 16. Results obtained at 100 MHz. The measurement error for the noise is less than 2% and the error for the CCE values is negligible.

We can see that the CCE is better in the first MIMOSA8 version. MIMOSA16 only gives good results in FPN as it was implemented in the non-digital OPTO 350 nm process. Uncompleted charge collection can be accounted for by slow diffusion [41] (diffusion is however, always a slow process compared to drift). For a squared micrometer and the first pixel layer the maximum hit rate is $0.07/10^6 = 7 \times 10^8$ per bunch crossing. For the ILC at 500 GeV the pulse repletion rate is 5 Hz with 2625 bunches per pulse. This leads to 13125 bunch crossings per second in average. This gives a hit rate of $10^{-3}$ per second in the inner layer and $3 \times 10^{-5}$ per second in the third layer. Note that with a 25 micrometre squared array the average hit rate in the inner layer would be 0.625 per second, which is relatively high. Note that with that configuration the length of the pulse is 1 ms and there is a dead time of 199 ms. This leads to and enhancement factor of 200 approximately.

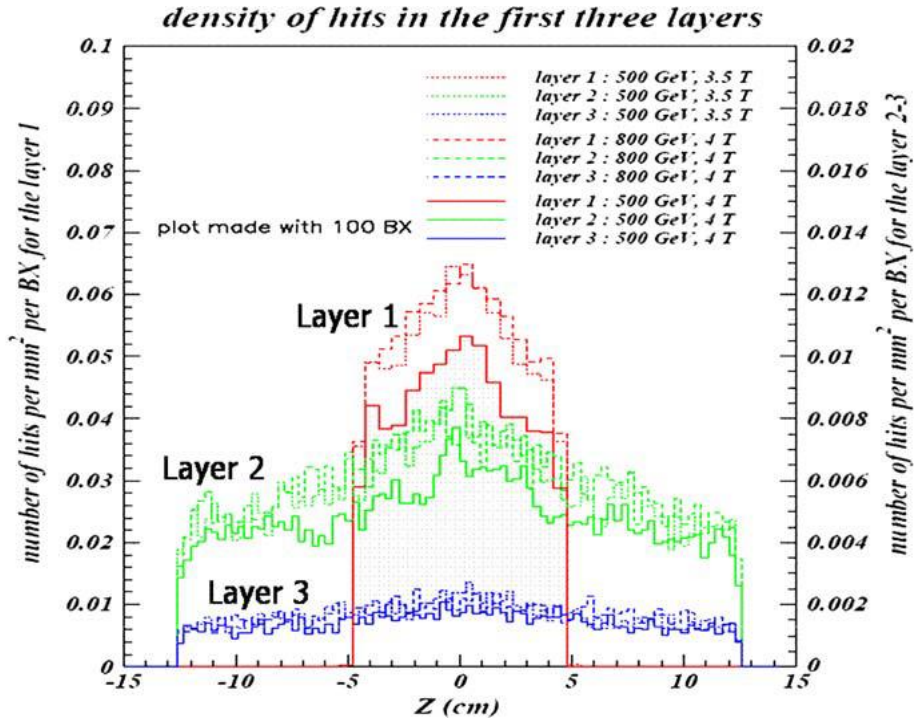

Fig.15: number of hits per bunch crossing/ squared mm$^2$ in three layers plotted versus the Z (incident beam) directions for the ILC [60]

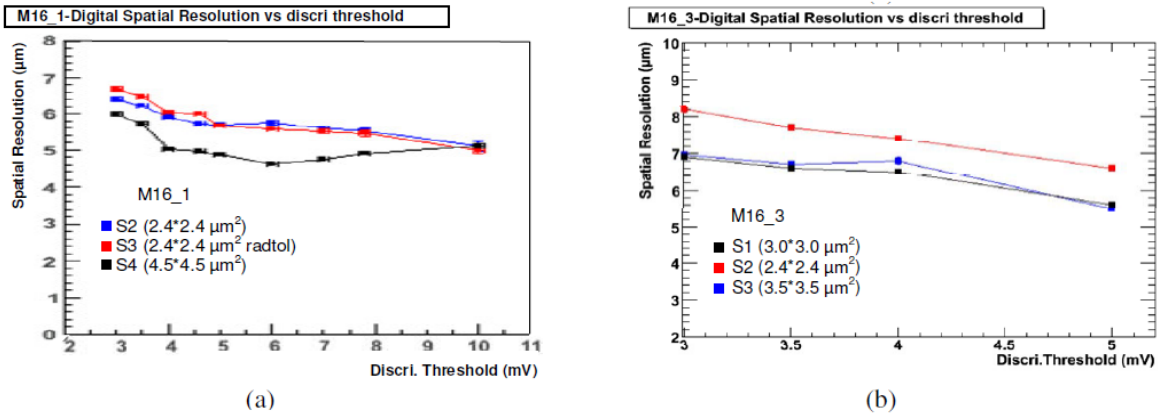

Fig.16: Results of the beam test, spatial resolution measured on the digital outputs on the MIMOSA16 chip. (CERN Beam tests ) (left 14 micrometer epi-substrate, right lightly doped non-epi substrate). The sizes are that of the diodes [35].

### 2.3.3. Experimental results: The problem of radiation effects

The radiation effects are key to obtain functional detectors at least in the inner part of the detecting system. We have to find out the a way to have a tolerance to a moderate value of 1MeV equivalent neutron irradiation and to ionizing irradiation. As these two techniques are good enough to explore the different aspects of radiation effects. The neutron induced crystal defects either point or extended [65]and the ionizing radiation charges the defects or polaron-states mainly in oxides [23]. For the MIMOSA8 no particular layout was used to obtain a radiation tolerant diode or transistor. An insulated guard ring is usually put between the p and the n electrode to deplete the surface and reduce the surface inversion layer. We have made the irradiation at CERI Orleans. The neutrons were spallation neutrons with a peak energy at around 14 MeV. The dosimetry was done with activated nickel. We have demonstrated that, the Charge Collection Efficiency, which is an important parameter, drops greatly above $10^{13}$ $n_{eq}$/cm$^2$. This is a disappointing result and showed at the time I had obtained it that this technology could not be used on hadrons machines. The pedestals, the pedestals dispersion (FPN) spread the temporal noise degrade significantly in this fluence range.

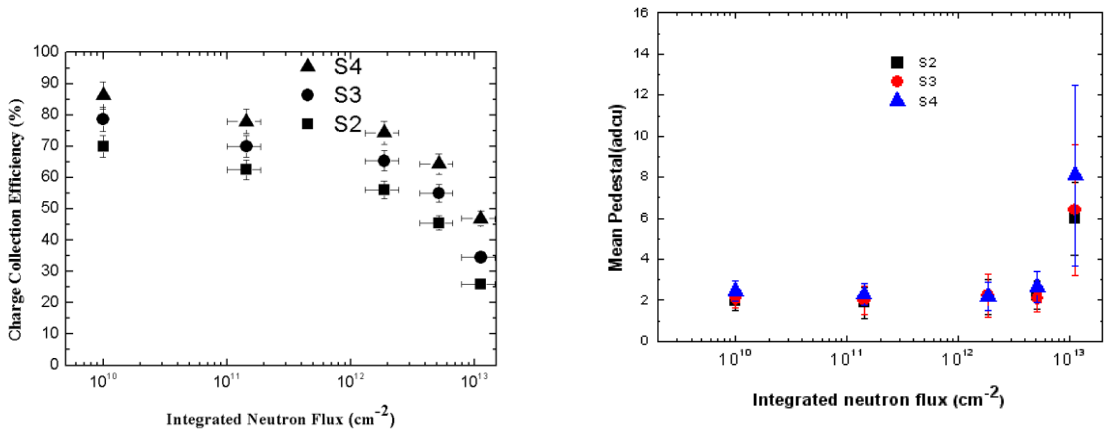

Fig.17: Charge collection efficiency for a standard clocking frequency versus neutron integrated flux. The neutrons were obtained from the CERI facility at Orleans France [38] [37] .

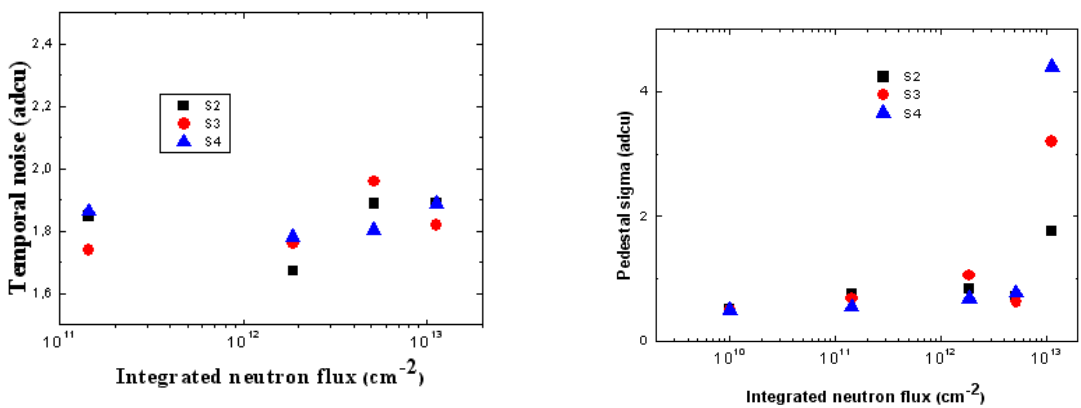

Fig.18: Temporal noise and FPN (dispersion of the pedestals) as a function of neutron flux [38] [37].

We can make a stop at this problem that is critical. First, the drop in CCE can be also accounted to the lack of fast carrier drift inside the pixel. As the concentration of carriers increases so does the capture rate of these carriers, this means that the traps fill rapidly and then no collection or only weak signal can be expected. I will

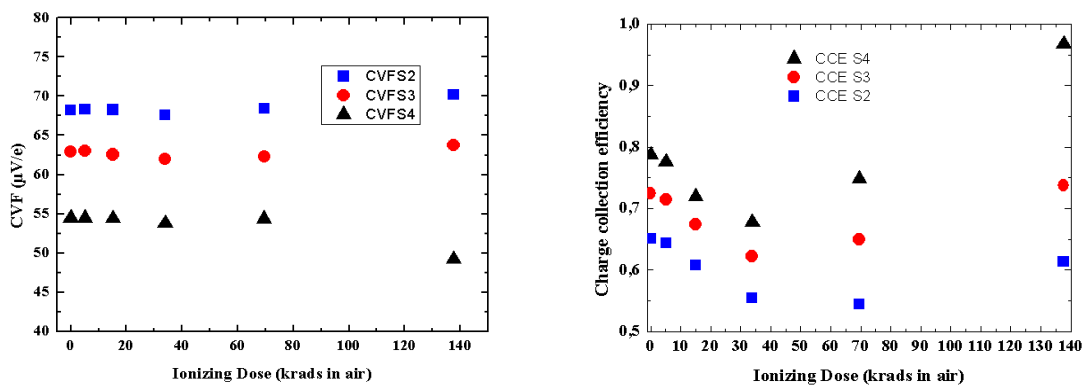

Fig19: CCE and CVF for the pixel-arrays irradiated with gamma ionizing radiation [38] [37].

develop this further on in the next paragraph.

The effects of ionizing radiation were dramatic but this can be accounted to a non-radiation hard design of the circuit and can be fixed (this had been done since on further chips). The surface and interfaces (Si/SiO2) can be held responsible for this behaviour due to the positive charging of the oxides as always observed.

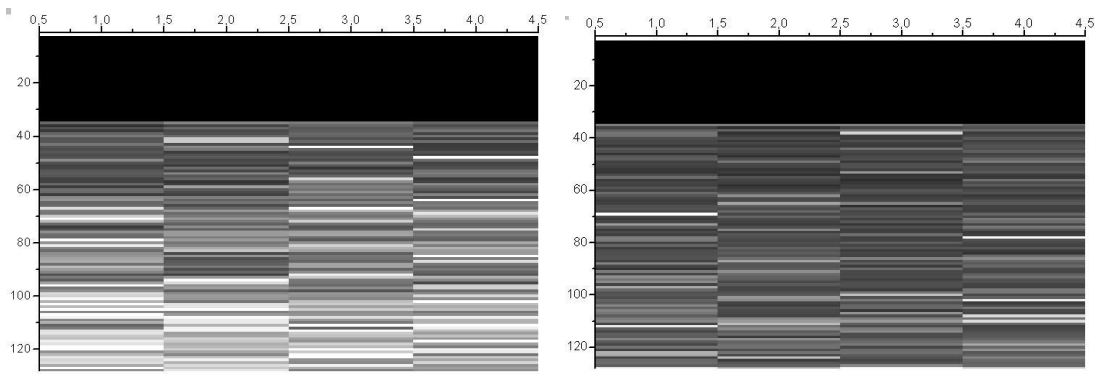

Fig.20: Fixed pattern noise on neutron irradiated pixels. The FPN is due to a non-uniform defect concentration. Each rectangular is a pixel [27] [65] [38].

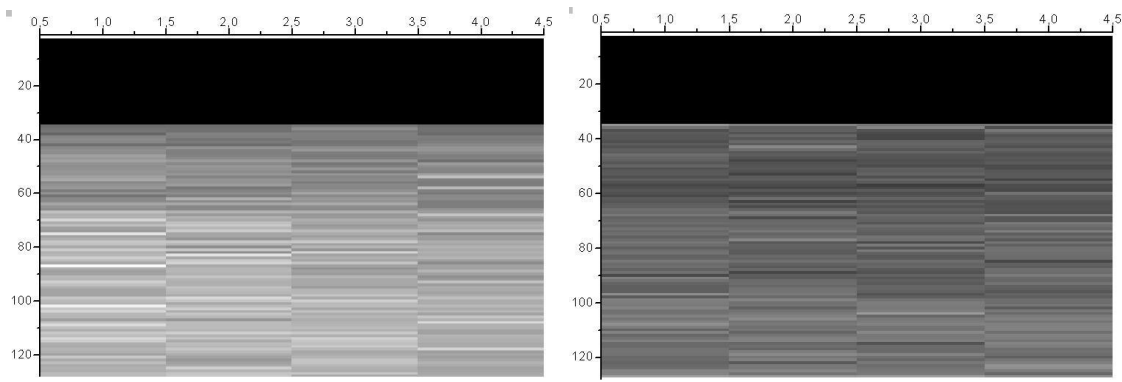

Fig.21: FPN of a gammy ray exposed pixel array. This is clearly no marked difference from one pixel to another. Irradiation dose 140 krads. No structural defects are created, only oxide charge is modified [27].

In a further analysis step, we have plot an image of the pedestals (FPN) values for first a neutron irradiated array and second for a 137 krads gamma irradiated array. As we have already demonstrated neutron irradiation induces a cascade of atomic displacements that are localized on a limited volume. Room temperature annealing or irradiation does mean that these primary defects can move and transform but experimental results from [65] showed that remain confined in a volume that is lower than 10 cubic microns

Fig. 20: shows that some pixels are more affected than they neighbours by neutron irradiation. This is consistent with the model already introduced. The non-homogeneous nature of the defect distribution with pixel with a lateral dimension of 25 microns is another demonstration of this phenomenon. This is a direct proof of the micron size of the defect rich zone. Fig.21: shows contrarily to the neutron irradiation that the ionizing effect are more homogeneous. This is not due to less contrasting. In this case no bulk effects are expected an all the FPN is due to the readout, not the detection cell which is a bulk device.

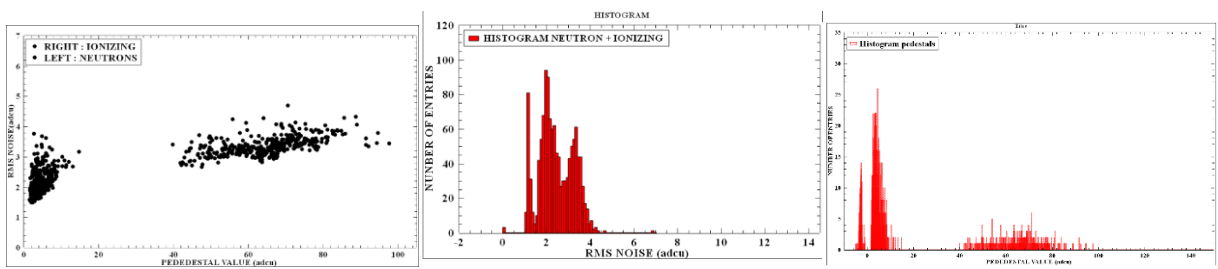

Fig.22: Temporal noise versus offset (pedestal), neutron on the left side and ionizing right side. The others are histograms for the rms noise and the pedestals [27] [37] [38] .

We can explore other consequences of the neutron/ionizing irradiation. The left figure shows a dependence of the temporal noise of each pixel with the pedestal (offset). The degraded pixels have a higher temporal noise because they suffer from the defects introduced by neutron irradiation Some G-R noise is generation in the PIN structure of the pixel. The dependence with the pedestals in strong and seemly linear but the pedestal remains at a reasonable value, attesting a limited threshold voltage shift. On the opposite, the 140 krad irradiated pixels exhibit a strong pedestal value and a spatial spread, which is due to the NMOS threshold voltage shift and distribution. The dependence of the temporal noise with the pedestal is not very strong but seemingly linear. In this case, the noise should be induced by the degradation of the oxides and oxides/silicon interfaces, which can induce G-R noise. The histograms on the right part of the figure show for the RMS noise distribution; this is possible to separate to the different contributions (neutron/ionizing). For the pedestal distribution, the neutron/ionizing can be easily separated, with the contribution of irradiated pixels centred on zero.

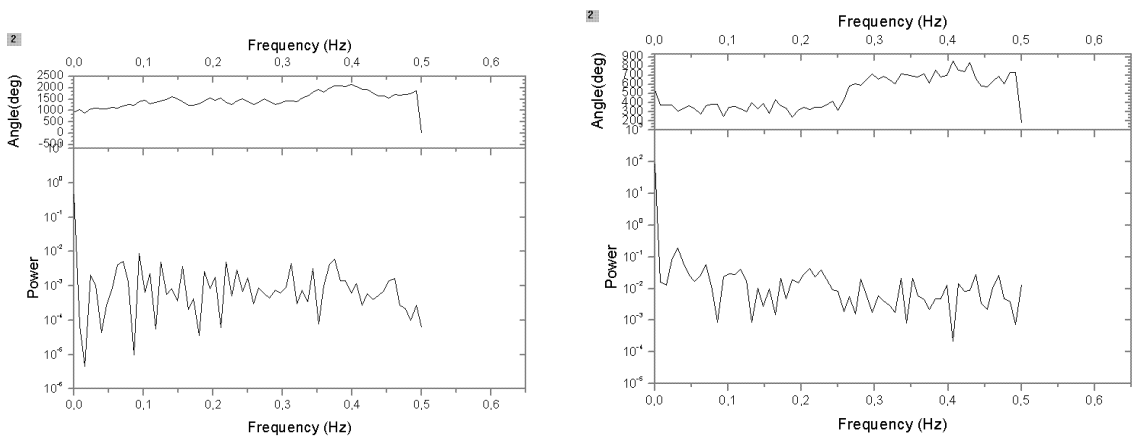

Fig. 23: Power spectral density of the FPN on the neutron-irradiated array, compared with that of the gamma ray exposed one. The frequency is indicated in Hz (this is a consequence of the software). Indeed, it is a spatial frequency [27].

In addition to the previous analysis, we have treated the FPN noise with signal processing methods. The Power spectral density of the neutron-irradiated array is higher than that of the ionizing-irradiated one. This shows that the neutron irradiated contain a signal more important than the ionizing one for the same spatial frequencies. The ionizing array has a constant (frequency=0) contribution, which can be clearly seen on the plots.

I can conclude this paragraph with the following remarks. First, the CMOS sensors with a pixel pitch of the order of 25 microns are very sensitive to displacement-induced defects resulting in the creation of deep defects in the sensitive volume of the detector. This is the most difficult problem that can only be solved by studying the pixel operation principle. Second ionizing irradiation induces threshold voltage shift and drop in the CCE that can be mitigated by using either:

      a) Radiation hardness design rules in the layout of the pixel (this was made in most designs)
      b) Radiation hard technologies such as SOI or more recently FDSOI

The standard CMOS sensors at the date of 2007-2010 was not able to cope with the constraints of hadron colliders but it was for for the ILC detectors constraints. This was mainly due to the operation mode of the pixels, which was based on diffusion and not on the drift of carriers in the sensitive area. The region below the surface is not depleted as it is in the case of hybrid pixels for instance in which an high reverse voltage is applied on a PIN structure made on lightly doped silicon.

## 2.3.4. Simulation of pixel structures

What is clear from the previous studies is the lack of appropriate model to describe the pixel and then to derive all its properties from it. The simulation tools are very effective to simulate the electronic readout and can be reliably used. The same can be said of the particle transport codes such as GEANT4 but the simulation codes for semiconductor devices were not until recently used in pixel or detector design. Many groups have started simulation by the years 2010 with simulations codes usually known under the TCAD acronym (Technological Computing Aided Design).

There are two ways to proceed to improve the pixel is to simulate its behaviour. I have made this to the radiation effects, which are the most detrimental to CMOS sensors.

The step toward an improvement of the pixel w;r;t; the neutron induced defects is to first have a good model to describe the transport in the structure. To make the simulation simpler one can eliminate the readout from the simulation script, as it does not affect the results.

Most of the results shown here were published in 2009 [39]. I analyse the most important aspects of them.

I have used a Silvaco-ATLAS software to evaluate the behaviour of the silicon pixel. More importantly, the deep defects must be introduce in the simulation code with appropriate value. In fact it was found that the simulation results were weakly depend on the energy levels but more on the capture cross sections. This comes from the effect described in section &1. We have used the deep level from Table 3 and the structure from Fig 24 right to obtain the results in Fig.24 left with 3 active layer thicknesses. The structure is hit by a charged particle track that we assume generate electron-hole pairs along it. Three different thicknesses were studied.

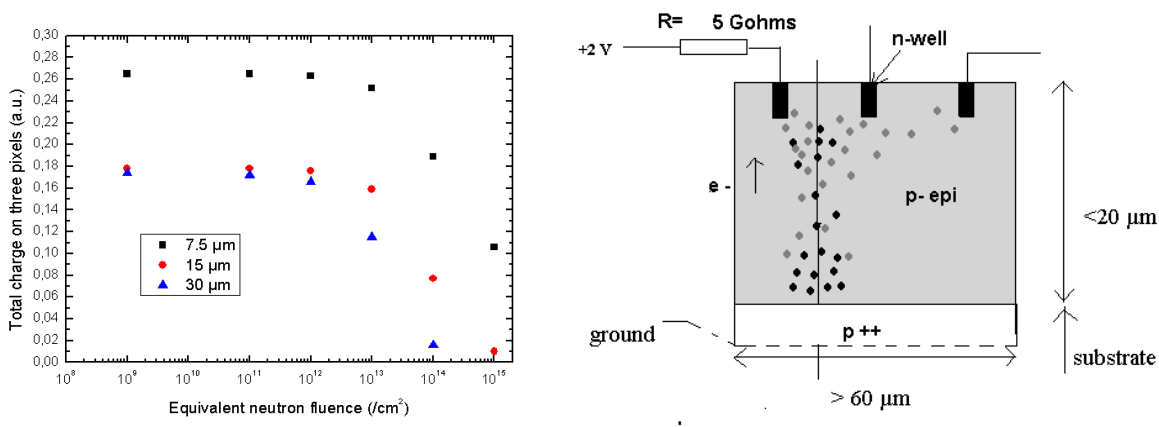

Fig.24: The computed charge collection efficiency was obtained with a simulated structure partially depleted in order to take into consideration the carrier diffusion [39]. The deep levels used for the simulation are in Table 3 .

Table 3: List of neutron induced deep levels used into the simulation. The capture cross section need are enhanced to take into consideration local disorder effects

| Traps considered | Capture cross sections | Introduction rate | Reference |
|---|---|---|---|
| $E_c$- 0.46 eV | Equivalent electron cross section $\sigma_e=10^{-13}$ cm$^2$ | 1 cm$^{-1}$ | FLE,MOL |
| $E_c$- 0.18 eV | Equivalent electron cross section $\sigma_e=10^{-13}$ cm$^2$ | 1 cm$^{-1}$ | FRE |
| $E_c$- 0.25 eV | Equivalent electron cross section $\sigma_e=10^{-13}$ cm$^2$ | 0.5 cm$^{-1}$ | FLE,MOL |
| $E_v$+0.36 eV | Equivalent hole cross section $\sigma_p=10^{-13}$ cm$^2$ | 1 cm$^{-1}$ | ERE |

What I obtain using a low voltage bias is a plot of the CCE that roughly fits the experimental one in Fig.25 with the effects of active layer (=detecting layer) thickness clearly appearing.

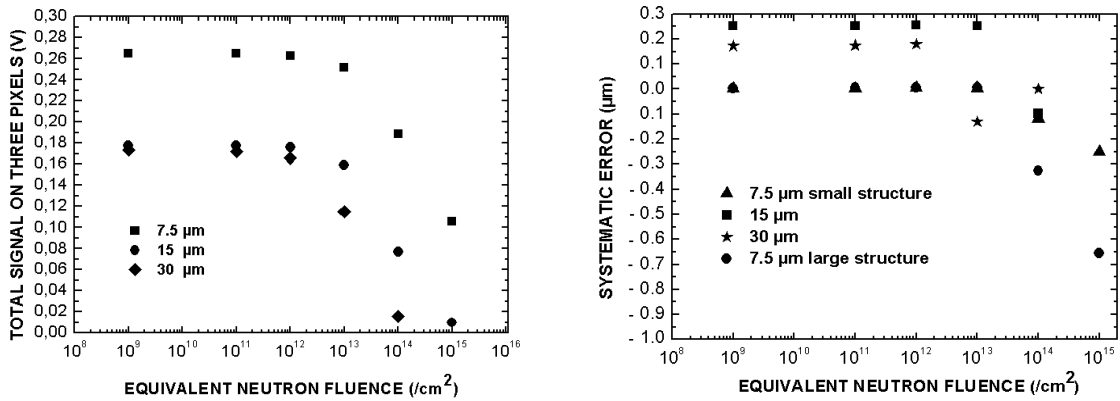

Fig.25: Neutron effects: signal and systematic errors on the hit position. The active layer thickness is set from 7.5 microns, 15 microns and 30 microns. A larger structure was used for the simulation of the 7.5-micron ALP (Active Layer Thickness) [39].

This also means that reducing the active layer thickness is a way to enhance radiation hardness. Nevertheless, in the case we have studied the effect is low. The doping level in the active layer is set to: $2 \times 10^{15}$ cm$^{-3}$. We find that the best way to increase the radiation hardening is to lower the doping level to get a depleted sensitive zone. That has been the trend since then

In the Fig 25 the effect of a reduced doping level on the signal is strong. The reduction of the net doping enhances the signal magnitude by a factor of four to five. This clearly demonstrate that the use of a depleted structure enhances charge collection and that potentially hard devices can should be based on the use of depleted structures small structures.

The other point stems from Fig.26 right. The effect of trap density on the total charge collected remains low as long as the values are below $10^{13}$ /cm$^2$. This limits the use of these first structures to low neutron fluences. When the defect density is very low the dependence of the total signal on active layer thickness is very weak.

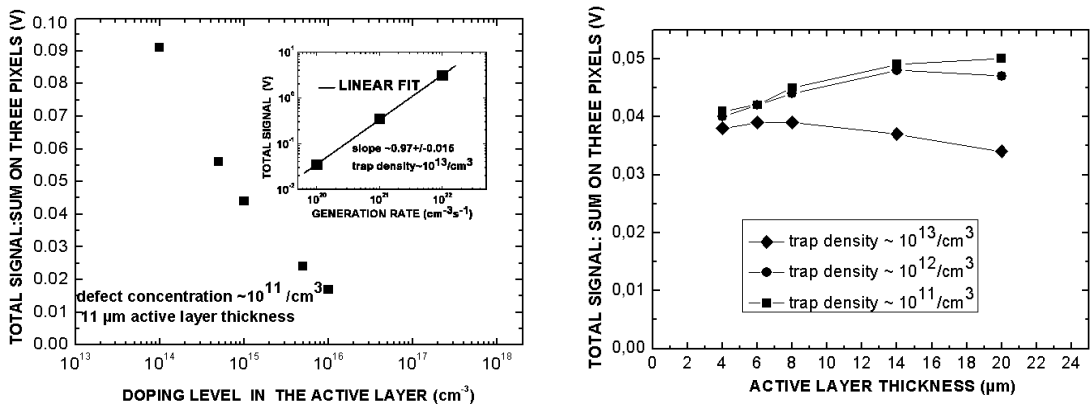

Fig.26: Simulated response of a three-pixel array with respect to doping level and trap density. It is clear that low doping levels lead to higher signals [39].

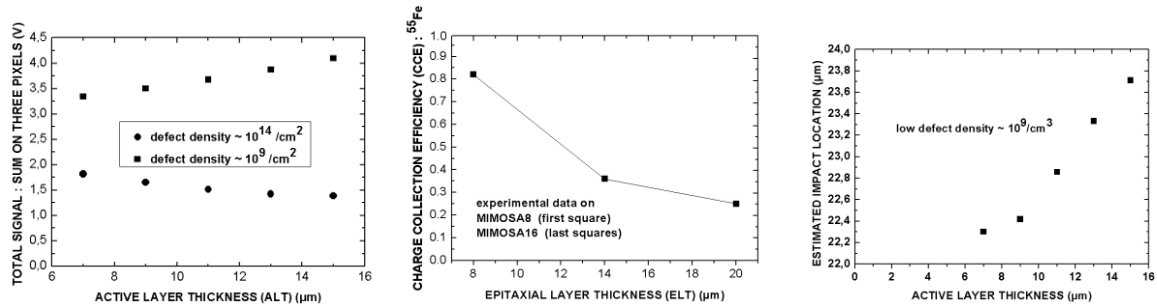

Fig.27: Estimated signal on the three pixels as a function of the active layer thickness (experiment and simulation) [39].

The Fig.26-27 show that the signal diminishes when the ALP (Active Layer Thickness) reduces high defects density and the opposite is true at low defect density. The reconstructed impact location is also thickness dependent. Experimental results with the ALP set to the epi-layer show that the signal behaviour is also consistent with the simulations. I have used device simulation to evaluate the sensitivity of the point of impact reconstruction (assuming vertical tracks) on the neutron irradiation (using the introduction rate of Table 3). I have used a COG method. The error (systematic) is increases as the irradiation fluence increases. The effect is less marked on thin in depth and small (in lateral dimension). This shows that the way to improve the performances is to use smaller structures than the ones used until now.

The following comments and conclusions can be made following this study. First, it seems a reasonable assumption to use simulation in the design and characterization of pixels as far as the electrical model is reliable and that reasonable parameters are introduced into the simulation script. In fact the structure that are simulated in pixel or semiconductor detector design are far simpler structures than the counterpart in microelectronics or nanostructures. MOS structures or bipolar utilize more transport models and parameters than then simple PIN structures. The second comment is that in many cases the separation between readout devices and sensitive device can be made in this case. This is a necessary condition in the case of pixels comprising some extra transistors devices (a few units to a few tens of units) because this reduces simulation time to reasonable values. One can easily make the simulation on a Lap/Desktop. This is not the case for pixel structures, I will introduce in the following sections. The TCAD simulations codes can then be used for technological evaluation of pixel structures. These are device (semiconductor/insulator/metal) material structures. Process development can be simulated (with restrictions) with these tools. I have mainly used them for feasibility study of certain structures not fabricated and not in many aspects studied.

## 2.4. The second step forward: other structures studied

Having taken into consideration the shortcomings of CMOS pixels detector, I have pursued my research in this field with three objectives.

1. First design a pixel with much improved spatial resolution by downscaling
2. Second : drastic radiation hardness improvements are sought
3. Use of device simulation is the preferred tool for this purpose

Following this prerequisite, I first applied these to the design and the characterization of a pixel based on a germanium structure. Yang Yuchao from the Toulouse M2 in nanoscience. made some of the work. I used the background of knowledge in Germanium neutron induced defect to get an improved picture of the evolution of defects as a function of time taking into consideration the possible thermal annealing at next to room temperature. Of course these are extrapolated from moderately high temperature data but for the sake of demonstration I deduce a first order annealing law that can be used for evaluation purposes. The effect of the extrapolated anneal is as expected. Using signal theory or other calculation the effect is that the defect concentration saturates after a time duration of a few hundredths of thousands of seconds. This corresponds to a 3-year duration. This is a relatively high value with respect to detector utilisation. The figures here are for germanium that has low annealing temperature with respect to its secondary point defects introduced after room temperature neutron irradiation. For silicon, the deep defects have a much higher annealing temperature, which render this mathematical procedure much more uncertain.

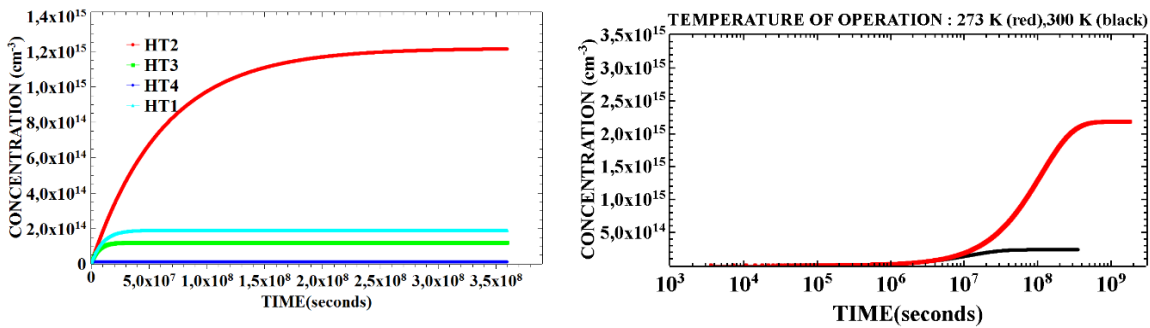

Fig.28: Case of radiation potential tolerance in germanium: the annealing process derived from a first order process at moderate temperature the red/right is at 273 K and the black is at 300 K. Fluence and time scale, neutron flux. [66]

To test a possible pixel structure we have here designed an Avalanche Photodiode structure with three doped layers zones made with Ge. The total thickness of the structure is 3 micrometres' which makes the number of primary electron-hole pairs generated by Minimum Ionizing Particles potentially less than 240 (or closer to 100 if we consider the slightly p-doped zone of $10^{14}$ cm$^{-3}$).

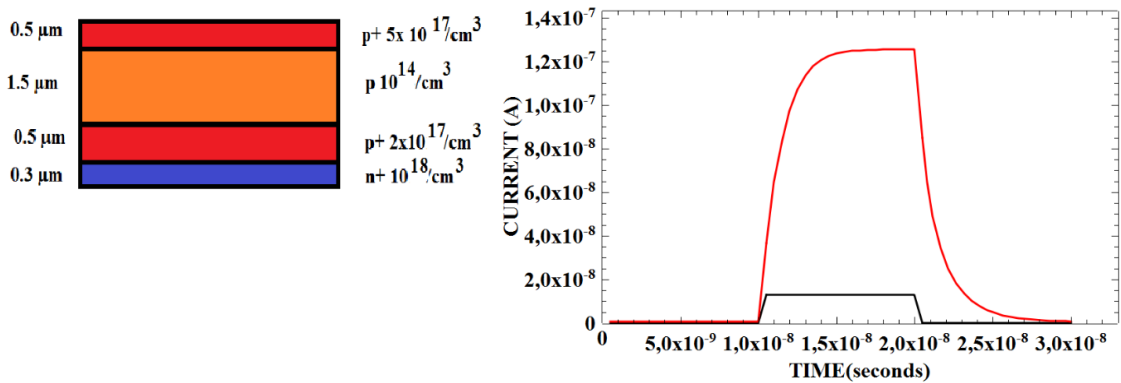

Fig.29: structure used for simulations of s Ge APD. Plot of the response of the APD versus time for a Minimum Ionizing Particle Track in avalanche mode. The reverse voltage was set to ~9 Volts with dimensions being 1 x 1 x 3 μm (3μm thickness). The structure was a p+ip+n+ with the amplifying zone being the middle p+ zone. The black curve corresponds to the excitation curve (the MIP). [66]

Signal amplification is made with high electric field induce by the p+n+ junction at the bottom of the structure (0.5 micrometers in thickness each). The device simulation of the structure is made using ATLAS-Silvaco device simulator and the result is shown in Fig. 29. The time response of the structure to an e-h pulse is of the order of 5 ns which can seem important for the structure studied. Moreover, the structure does not latch-up. In this case, that means that it does not short circuit due to the injection of holes from the p+ doped region or the injection of electrons through the structure. In the case of silicon APD, a resistor is needed to damp this effect. The current flow through the resistor lowers the voltage at the APD electrodes and hence switch it off. Most results for this structure were publish in [66]. Moreover, the critical fields for avalanche mode. As an outcome of this research, it is clear that device simulation of the structure is a step that is a prerequisite for device design. The software used here is SPISCES and ATLAS and in addition SUPREM4 and ATHENA for technological design.

### 2.4.1. Particle transport through silicon: simulation

We have concluded that device simulation should be used for pixel detector development. This means that the effects of particle tracks through the structure should be known with some accuracy. This is only possible with the help of high energy particle simulation codes such as GEANT (GEometry ANd Tracking).

Table 4: Simulation results made with GEANT4 on silicon pixels (or their equivalent in material budget). V. Kumar (as an Ecole Mines Nantes student) made these simulations [67].

| Pixel Size (µm²) X Thickness (µm) | $E_{avg}$ (keV) | $E_{MPV}$ (keV) | Ratio avg (%) ($E_N$/$E_p$)* | Efficiency ($E_{Th}$= 500eV) |
|---|---|---|---|---|
| 1 * 1 * 10 | 3.28 | 2.17 | 4.74 | 99.3 |
| 1 * 1 * 20 | 6.76 | 4.89 | 4.9 | 99.82 |
| 1 * 1 * 30 | 10.86 | 7.84 | - | >99 |
| 1 * 1 * 40 | 14.36 | 9.68 | - | >99 |
| 1 * 1 * 50 | 17.98 | 13.21 | - | >99 |
| 10 * 10 * 10 | 3.26 | 2.15 | 0.994  1 | 99.9 |
| 10 * 10 * 50 | 16.68 | 13.17 | 2.08 | >99 |
| 10 * 10 * 100 | 34.76 | 24.3 | 2.35 | >99 |

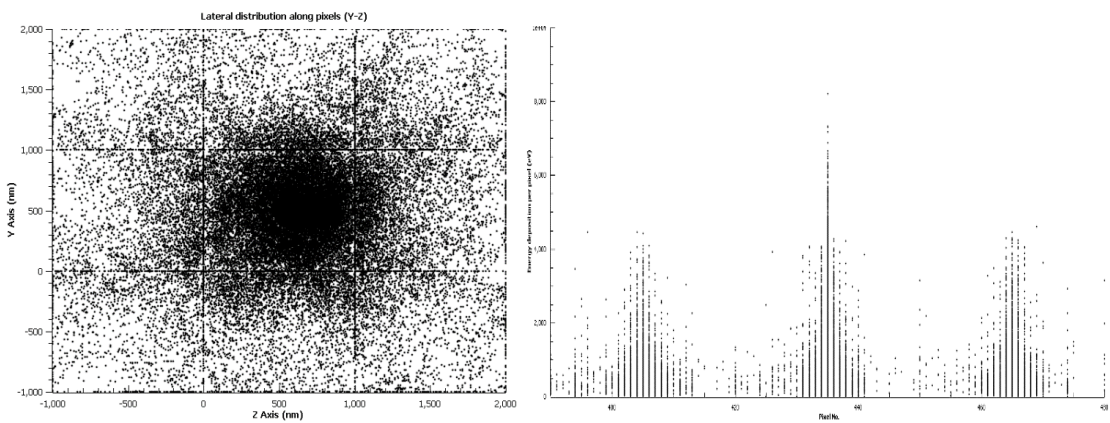

Fig. 30: Simulation of the passage of charge particle through a silicon thickness of 50 micrometer [67]

We may conclude that the pixel size can be reduced to 1000 nm in pitch if the thickness of the silicon substrate is set to less than approximately 50 micrometers. The hit spread is here shown as less than one micron. See above Fig. 30. One conclusion is that a Gaussian law can approximate the distribution of the hits. In other terms, the distribution is not uniform and the plots above seem to indicate a normal or Lorentzian distribution. In spite of the Lorentzian shape, which fits with the scattering theory, a Gaussian simulation seems adequate

### 2.4.2. Noise in pixels: computation and simulation

We have seen that the predicted detection efficiency remains high enough ( > 98 %) with pixels thicknesses down to 10 micrometers. This with a relatively high threshold > 500 eV corresponds to 500/3.6 =140 electron-hole pairs. That also means that the Signal to Noise ratio can be of the order of 10 if we assume

ENC (Equivalent Noise Charge) of 14 electrons. We expect to reduce this noise further on by downscaling the device. The formula used in paper indicates that the ENC strongly reduces with the L and W [56] [57].

$$ENC^2 = ENC^2_{series} + ENC^2_{//} \quad (1)$$

Ci is the input capacitance and $g_m$ the transconductance the other terms are defined below.

We hence define the parallel and series noise and compute analytically the two terms:

$$ENC^2_{series} = 2k\theta \, (1+\frac{Ci}{C_{gs}})^2 \, \frac{CiC_{gs}}{B} (1-e^{-2T_{obs}g_m\frac{Ci}{BC_{gs}}})(5)$$

$$ENC^2_{//} = qI_{leak} \, \beta^2 \, \frac{Ci}{Ag_m}(1-e^{-2T_{obs}g_m\frac{Ci}{A}}) + 2qI_{leak} T_{obs} \quad (9)$$

In addition there is a low frequency 1/f term (flicker noise)

$$ENC^2_{1/ft} = \frac{\pi A_\omega T_{obs}}{2B^2 \alpha}\left(1+\frac{Ci}{C_{gs}}\right)^2 Ci^2 \, (13)$$

$\alpha = \frac{g_m Ci}{BC_{gs}}$ (11) is a cut-off frequency. $A = BC_{gs}$ (7) and $\beta = C_{gs} - \frac{A}{Ci}$ (8)

B ~ Ci +C  (there is an error in the expression of Ci) the last tem contains a C prefactor

with $Ci = C_{gd} + C_{diode}$

B=C+ $C_{gd}$ +C $C_{gd}$/$C_{gs}$ (3) and not:

$$B = C + Ci + \frac{Ci}{C_{gs}} \, (3)$$

C is the output load capacitance.

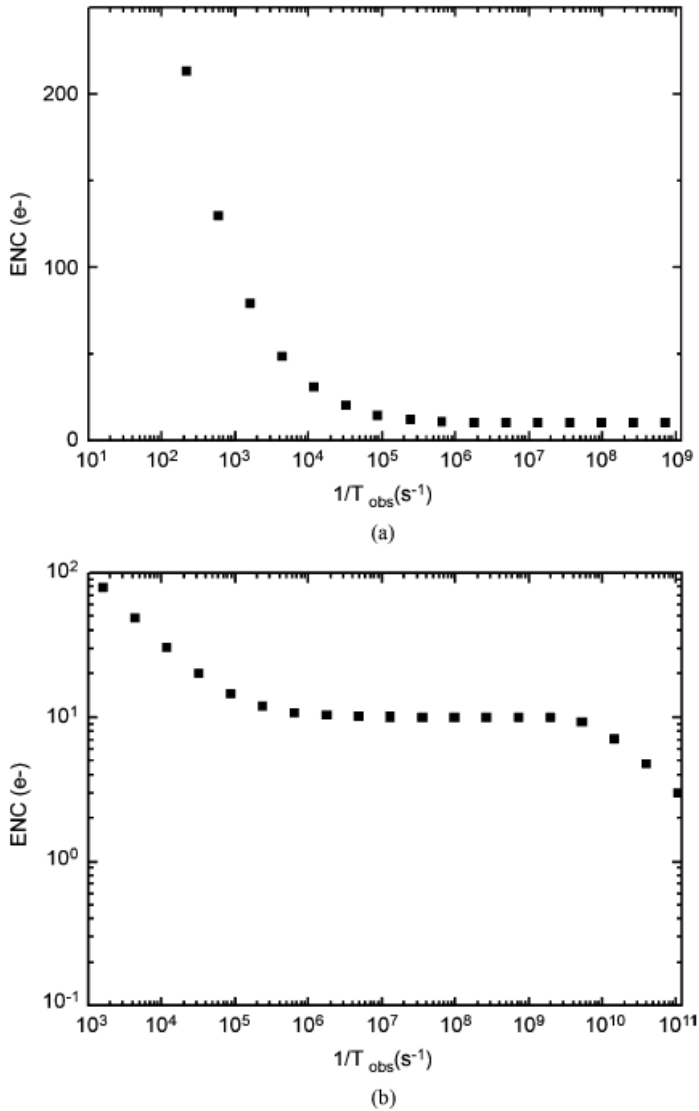

Fig. 2. Theoretical ENC versus reciprocal of the observation time, a plateau is reached above a value, which experimentally corresponds to a relatively low clocking frequency. (a) in linear-log scale (b) in log-log scale.

The ENC decreases when the capacitance (input and readout) reduces. For the series ENC, noise the equivalent noise charge is due to the readout circuit and should be inversely proportional to the capacitance both in the input and output. The ENC series noise is directly related to the Johnson/Nyquist noise of the NMOS channel. We can reduce it by downscaling. The ENC parallel is proportional to the square root of the leakage current. This is a Schottky noise and can be reduced by downscaling the structure, as the leakage current is proportional to the perimeter or the area of the device (for the surface effect such as the ionizing effects). The G-R current is proportional to the volume of the sensitive device (PIN structure) and hence to bulk effects such as the neutron irradiation (and NIEL). Decrease of the dimensions also have favorable effects on the capacitance and the trans-conductance and hence on the ENC parallel. The low frequency noise or called (Flicker Noise, flickering or fluctuation) is modeled as a ENC 1/f noise and behave like a series noise and is be less dependent on the device size, but can be reduced using high frequency sampling and transconductance increase. The next table summarizes the trends:

Table 5: influence of the device geometrical and operation characteristics on the noise of the 3T pixel.

| Parameter | Capacitance decrease | Sampling rate (1/t) increase | Trans-conductance (gm) increase | Temperature (K) decrease | Leakage current decrease |
|---|---|---|---|---|---|
| ENC Schottky | decrease | decrease | decreases | Not applicable | decreases |
| ENC Johnson | decrease | Decreases | Reaches a limit (kT/C) | Proportional decrease | Not applicable |
| ENC Flicker | decrease | decrease | decrease | Not applicable | Not applicable |

Note that the change in the geometrical parameters W (width of the elemental transistor) and L (length of the elemental transistor) have the following consequence.

   a) W decrease: capacitance decrease transconductance increase , possible sampling rate increase., possible leakage current decrease
   b) L decrease: capacitance decrease transconductance increase, possible sampling rate increase. possible leakage current decrease
   c) Temperature decrease: Johnson an Flicker Noise decrease , leakage current decrease

We can conclude from this discussion that a reduction in size is favorable to noise reduction.

### 2.4.3.  Global pixel design

We can now give a set of conditions if we want to have an optimal geometry with adequate data flow compatible with state of the art technology. We assume that no other limitations influence our model. Particularly we set aside the transport of high-energy charged particles into with the scatter in the silicon. A purely numerical and analytical estimation for the effects of downscaling is presented in the following table.

We have used the following parameters:

   a) Pixel size or pitch , lateral dimensions
   b) Number of hits per unit are and per seconds.

We then have the following results.

   a) Data flow (assuming one bit pixels)
   b) Number of pixels per identical array
   c) Address length in bits

Table 6 below summarizes the results.

Table 6: size of the pixels, area, number of hit per unit area, address length, data flow [67].

| Size (lateral dimensions) | Resolution (first order binary) | Area | Number of pixels Np in an array | Number of hits per unit area and per second | Address length in bits N=log(Np)/log(2) | Data flow in bits/second |
|---|---|---|---|---|---|---|
| 1x1 µm x µm | ~ 1 µm | 10 cm squared | $10^9$ | N | 30 | 30 x N |
| 10x1 µm x µm | ~ 3 µm | 10 cm squared | $10^8$ | N | 27 | 27 x N |
| 10x10 µm x µm | ~ 10µm | 10 cm squared | $10^7$ | N | 24 | 24 x N |

These figures show that choice of a 1 x 1 micrometre squared pixel is still a reasonable choice, compatible with the present technological nodes in data processing for instance [67].

### 2.4.4. A new pixel concept

I have pursued this research project by introducing a new pixel architect. Simply said, the pixel reduces to one single device. The device evolves from two distinct pixel architectures. First the DEPFET pixel that uses a buried gate underneath the channel for current flow modulation, it is a rather big device both in lateral size and depth. Second the CMOS sensor with a sensing photodiode and in its simple 3T biasing and readout devices. The evolution is shown in the next figure.

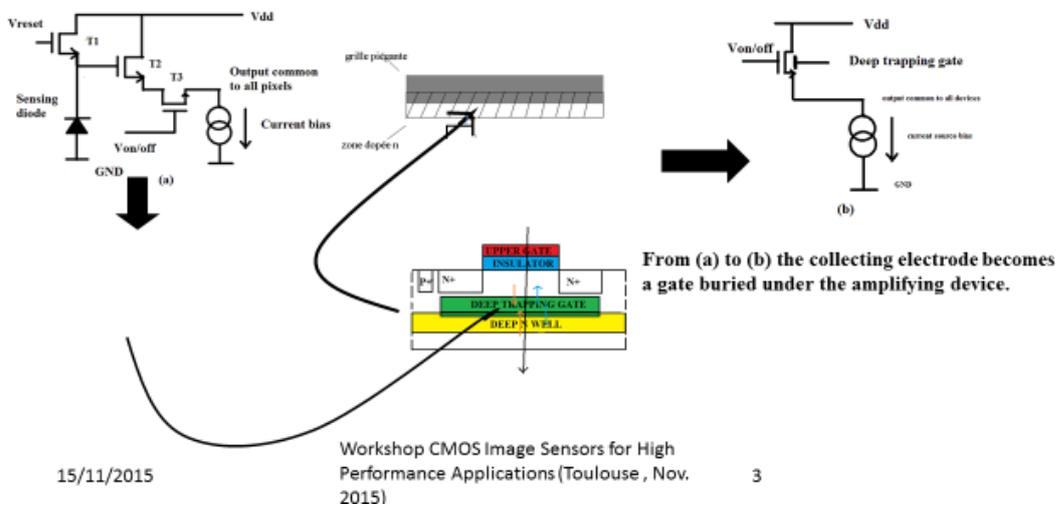

Fig.31: evolution from the 3T pixel sensor to the 1T buried gate pixel. The pixel is represented as being biased permanently [68] [69] [70].

In order to test the concept we have simply used a MOS structure on which we have added a buried gate composed with deep levels. The levels use for the test of functionality were substitutional Zn. This impurity is a double deep acceptor in silicon. The operational principle was studies in. The impinging particle give rise to electron hole pairs that trap in the deep-level doped zone for the hole and modify the potential variation through the structure with a buried gate getting more positive. The structure of the pixel is shown on Fig. 32, with the corresponding schematic.

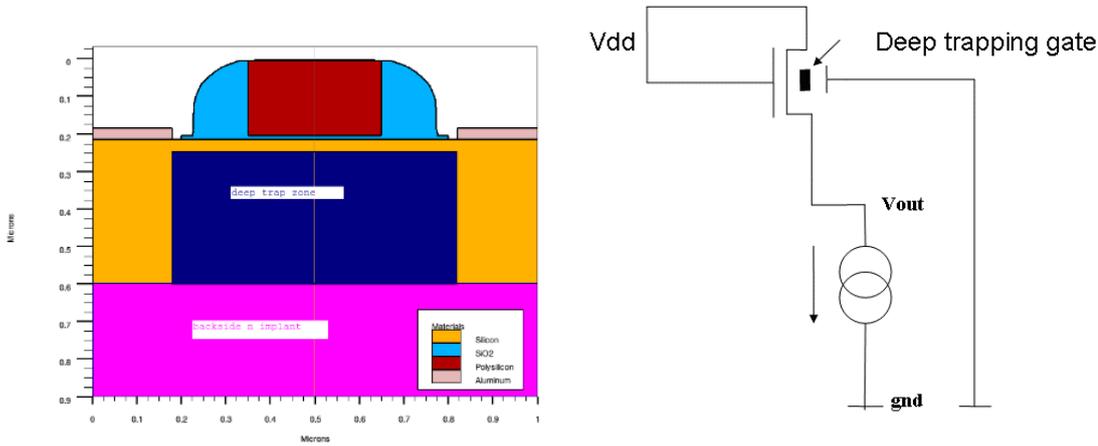

Fig.32: Structure of the TRAMOS used in the simulation code. First, the TRAMOS with trapping centres and second the proposed Quantum Well with Ge a material of choice see [42] [71].

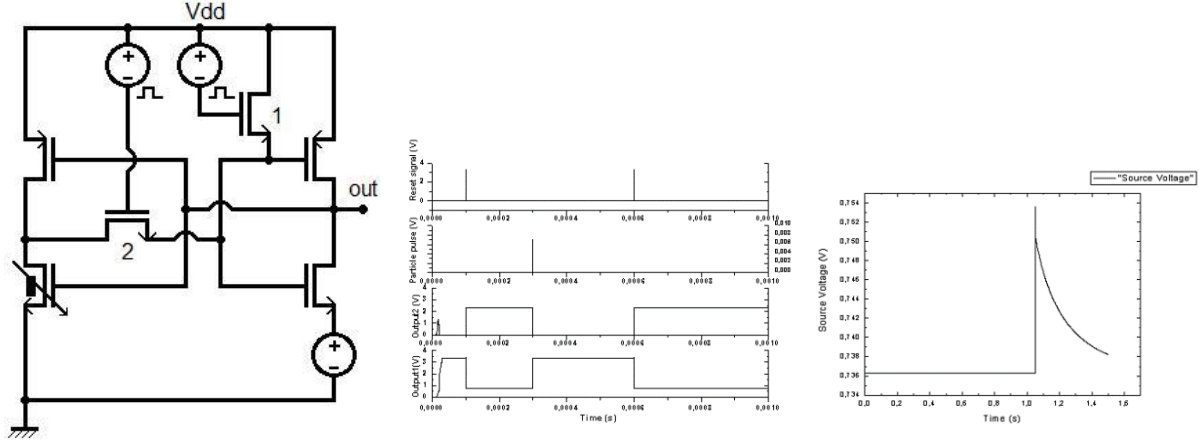

Fig.33: Schematics of the originally proposed latched pixel using the first design [42] [72].

Despite a number of drawbacks we shall discuss further on, the device can be introduced in a latch with a control logic, which may be, and effective way of design a binary pixel with a reduced number of devices. However, this design still requires some control and biasing lines. The structure proposed here and called TRAMOS (for TRApping MOS) has a number of drawbacks. First, it needs to be constantly biased to operate and second how the transport of holes in the trapping layer is not fully understood. For the operation, it needs a potential barrier that is made of a n+ deep –n-well at the bottom of the structure. In the figure below the structure is shown on the left with a deep n well and a trapping region. In this configuration below the trapping region is also a well for the holes in both cases, for a Ge quantum well too. This configuration requires the use of a highly doped n-well to be functional. The domain in which the holes can be collected is very limited, so that the S/N of such a device should be lower than standard CMOS sensor.

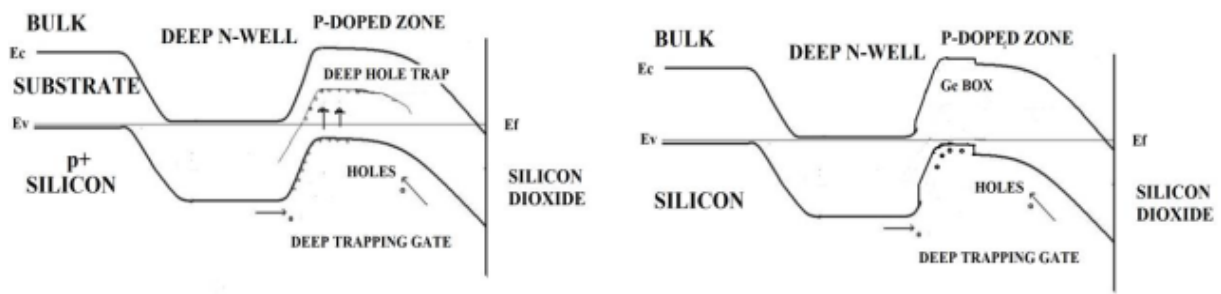

Fig.34: band diagram of the TRAMOS structure together with that of the Ge box structure [71] [70] [69]

To overcome these difficulties I have introduced a further concept where the deep n-well is not present and hence with a thickness of the active layer is of a few microns. This layer should be resistive silicon with a buried electrode located a few microns under the surface. This electrode would be necessary to screen the active zone from the bulk of the substrate. I shall describe how it possible to implant a layer deep in the bulk (6 microns below the surface).

The buried gate should have the following properties:

a) Ability to trap or to localize carriers
b) Ability to discriminate between carriers

The first proposed solution is to use a substitutional impurity able to trap holes only. With a Zn doped silicon layer, this is possible. However, this solution has many disadvantages. First Zn can diffuse during the processing of the device. It may act as a contaminant leading to dysfunctionalities. We have instigated the properties of the Zn implanted layers using a few experimental techniques such as SIMS, RBS, DLTS and Raman spectroscopy. Some of the results were published in [67] [71] [69] [70].

What stems from the measurements is that the un-annealed samples do show some instabilities. This is clear on the Schottky contacts made for the purpose of DLTS measurements. The variation of the I(V) characteristics with time clearly indicate that the implanted layer is not stable. DLTS measurements show that the levels due to Zn are metastable. The peak vanishes or is strongly quenched from one scan to the other. The results of upward scanning are different from the results of downwards scanning. Thermal Annealing would be a part of the solution at the expense of possible contamination. We may note that the principle of 1T pixel with doped buried layer was described in some earlier work but there was no following development. In this case, only a buried layer highly doped with shallow impurities was considered. We have then investigated another possible solution to circumvent the potential shortcomings of a highly doped layer with deep levels. The first is to obtain the same effect than with deep levels. The solution is to create a Quantum Well for one of the two carriers and second to have a fabrication process that has a reduced contamination effects.

For this purpose, the use of III-V semiconductors could be envisaged but that would involve a total change of substrate up to now (Note 2). In the present day technologies it is still difficult to make such a process compatible with a silicon CMOS process. The fabrication of MISFET is still not standard. The closest device is a MESFET.

We have proposed to use Ge as the material of choice to make hole quantum wells. Ge can be used in SiGe alloys [73]. The following table shows the results obtained prior this study on the SiGe layer in a silicon matrix.

Table 7: SiGe structures with their band offsets for different lattice configurations

| | Relaxed Si layer (Substrate) | Strained Si layer (tensile strained) (substrate) |
|---|---|---|
| Strained Ge, SiGe layer (compressive) | $\Delta VB \Rightarrow -0.66$-$0.74$ eV Quantum Well (valence band offset, well in the Ge layer) | ///// |
| Strained Ge, SiGe layer (compressive) | $\Delta VC = +0.00$ eV Quantum Barrier (conduction band offset, barrier in the Ge layer) | ///// |
| Strained Ge, SiGe layer | Eg=0.68 eV (gap) for the Ge layer | ///// |
| Relaxed Ge, SiGe layer | bulk properties | $\Delta VB \Rightarrow 0.66$-$0.74$ eV Quantum Well in the GeSi layer |
| Relaxed Ge, SiGe layer | | $\Delta VC = +0.24$eV Quantum Barrier in the GeSi, Ge layer |

We can conclude that when using a relaxed or compressively strained SiGe layer a valence band offset is created in the SiGe layer. This VB offset is a well for the holes. Additionally a barrier for electron in the conduction band arises when the SiGe is compressively strained. In the case of relaxed or stained layers, we should expect a VB well, which should act as an artificial trap with a sufficient number of states to make a few holes, trapped in the QW. Analysing the system in a classical approach leads to the structure here represented with a buried Ge layer. We have represented in Fig. 35 the structure used for simulations and the corresponding Si/Ge band diagram.

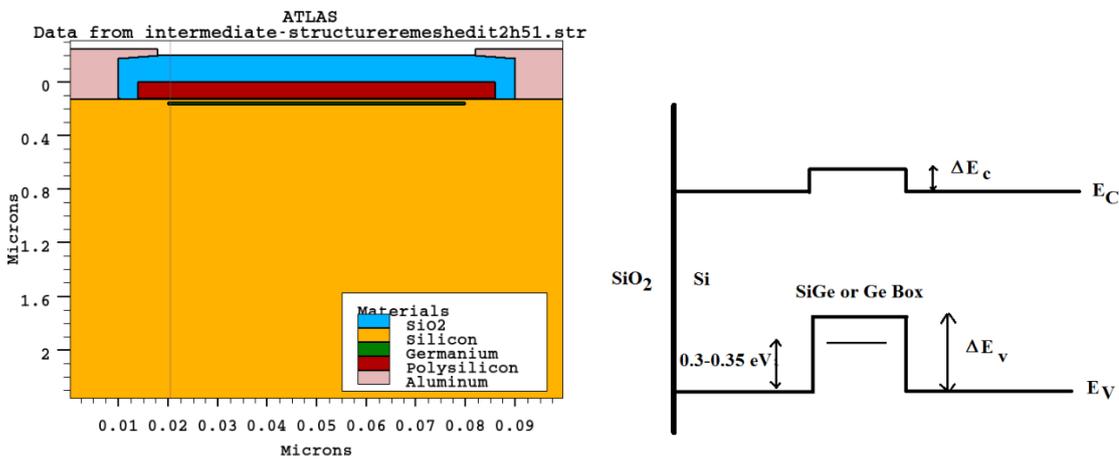

Fig.35: Quantum well with Ge materials, such as constrained and relaxed Ge. The band offsets are from ref. [73]

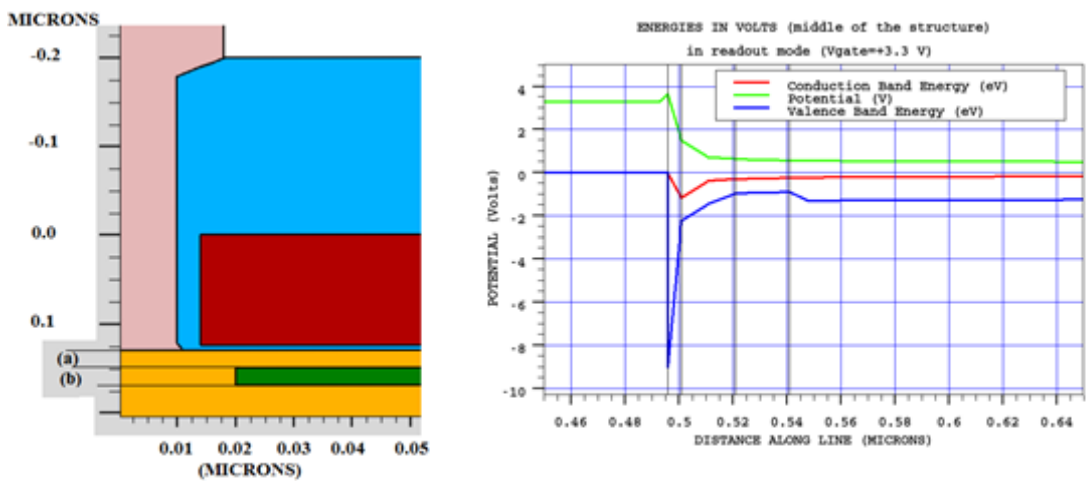

Fig.36: part of the structure, which is active for charge collection and current modulation, and corresponding band diagram the software used is Silvaco ATLAS.

The above-mentioned structure is the detailed structure and on the right the band diagram of the structure given by software simulations. We have done extensive simulations of the proposed device. The particle track is represented by the traversing line setting, with a number of electron-hole pair generated of 80 per track micron. The response to this was studied with quantum corrections (Note 3).

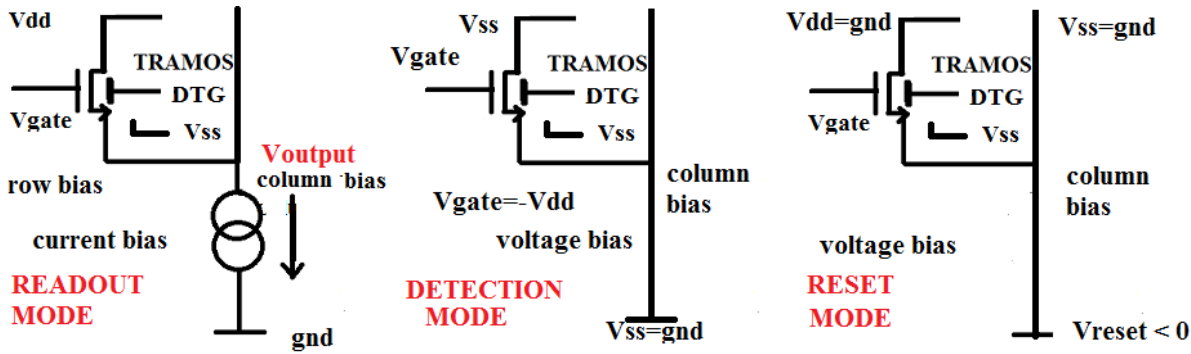

Fig.37: Operational principle of the QWELL pixel. One can observe similarities with the DEPFET principle. The evolution from CMOS pixel to the Quantum well structure is shown here. The pixel reduces to one single transistor. Biasing scheme for this device. The device can operate in detection, readout and reset mode. Another mode is the blind mode in which the pixel should be insensitive to charged particles [42].

The device has three operation modes represented in figure (Fig.37) we can summarize below:

a) The Detection Mode in which the Upper gate is negatively biased Vgate=Val <0 . In this mode, Vdd and Vss can be either grounded or kept at Vgate.In this mode the transistor is off
b) The Readout Mode: the upper gate is positively biased to switch the transistor in the On mode. The Drain is positively biased and the source is biased in current mode.This is the only mode where power is dissipated.
c) The reset mode in which the source is negatively biased and such is the case for the bulk and the drain grounded. An electron flow from the source is injected in the whole transistor and through the QW with holes recombining. The gate is positively biased. The substrate is grounded so that the electrons flow through the buried gate.
d) Another mode exists in which the Upper gate, Drain Source are grounded and so is the bulk, the holes in the QW should remain trapped and no increase due to ionizing particles should occur.

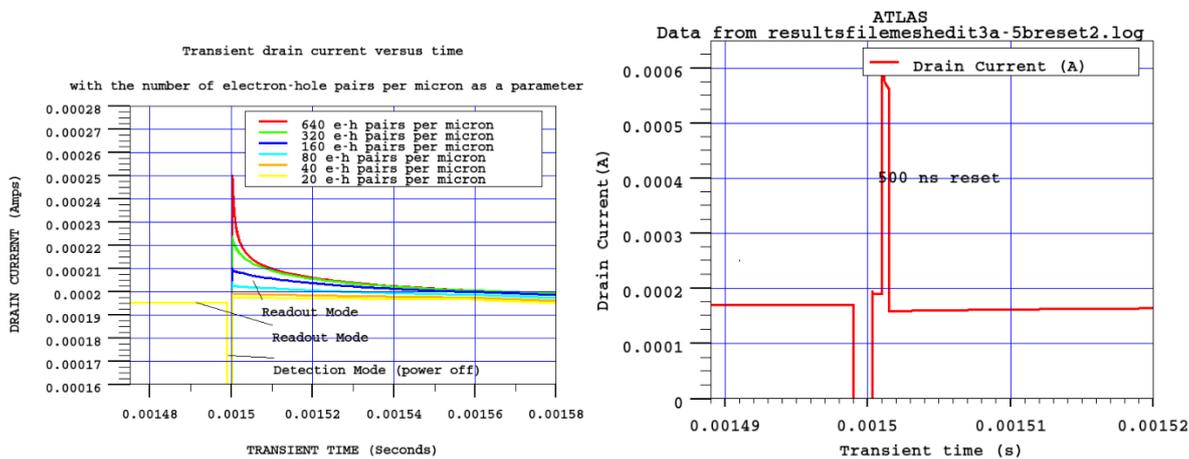

Fig. 38: simulation of the pixel in detection and readout modes if no hard reset is made the pixel self-resets after around a few tens on microseconds. Hard reset can be made by injecting electrons in the QWELL. Alternatively, by extracting holes from the well by biasing the source and drain positively or the substrate strongly negatively, the gate being unbiased or positively biased. A top bulk p+ contact would be very useful for this purpose [42].

The simulations results show that a good response is obtained for the pixel. A good sensitivity is obtained for a 10 micrometres thick structure. A significant signal is obtained for a 400 e/h pairs signal.

A conversion factor (CVF) of 15 nA/e (with respect to a signal obtained with no illumination) is obtained on a device of 1/0.1-µm aspect ratio with the source load of 5 k corresponding to a CVF of up to 100 µV/e (or hole in this case).

The simulations have shown that the proportion of 50% of Ge in the buried gate is necessary for good operation. With this in mind the device has a good linearity is obtained for signals down to less than 200 electrons. The behaviour is closer to a logarithmic dependence for stronger signals. Another simulation with the lifetime of the holes and electrons set at 100 ps show that the device is still operational with readout times up to more than 10 microseconds. The potential for radiation hardening against displacement-induced defects is established if we consider that, the lifetime is the essential parameter used in this simulation.

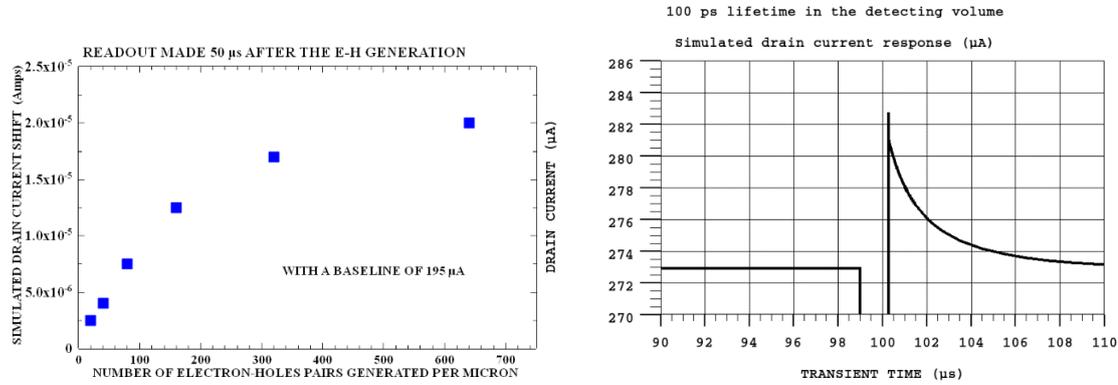

Fig.39: Simulation with a carrier lifetime set to low value in order to evaluate hardness potential [42].

### 2.4.5. Simulation techniques and physics

The structure proposed here is at a scale that means that careful simulations strategies are necessary. The semi-classical approach to semiconductor device behaviour fails is the dimensions of the device are too low (reference needed) and this is the case in nanoscale devices. In our case the dimension are close to the micron scale. This means that in a first simulation approach the transport can be treated in a semi-classical way. The drift-diffusion transport model is used to make the transistor mode simulations. Quasi-static characteristics are obtained this way. The use of quantum corrections is not necessary (A. Tenart report). The dimensions of the device and especially the scale of the buried gate imposes to check the effects of quantum confinement and effects on transport. As a limited number of models are implemented in the simulations software, we can use:

a) Schrödinger-Poisson model which is a recursive method to determine the Eigenfunctions. If the system is steady state this is accurate but is time consuming especially in 3D. It is not adapted to transient simulations for this reason, because it would need solving the coupled Schrodinger-Poisson equations at each step. This approach was one of the first used for quantum well modelling. This may be used in 3D at the expense of simulation time [74].
b) Density gradient model: this model is based on a transport equation derived from a quantum (effective) potential $\Lambda$.

$$\vec{J}_n = qD_n \nabla n - qn\mu_n \nabla(\psi - \Lambda) - \mu_n n(kT_L \nabla(\ln n_{ie}))$$

The so-called Wigner distribution function is used in this case defined as

$$W(x,p) = \frac{1}{\pi \hbar} \int_{-\infty}^{\infty} \langle x+y|\hat{\rho}|x-y\rangle e^{-2ipy/\hbar}\, dy,$$

$\rho$ is the density operator and $\hbar$ the reduced Planck constant

We can deduced moments of the distribution to determine the observables. The density operator is deduced from the canonical definition.

$$W(x,p) \overset{\text{def}}{=} \frac{1}{\pi\hbar} \int_{-\infty}^{\infty} \psi^*(x+y)\psi(x-y)e^{2ipy/\hbar}\,dy$$

This approach is based on the definition of a density matrix with field operators, (instead of wave functions) is the space representation. The time evolution of the system can also be calculated from the Schrodinger equation. The density operator is equal to : $<\phi^+(x)\,\phi^+(x)>$ which turns out to be dependent on time. In Simulation of Ultra-Small GaAs MESFET Using Quantum Moment Equations [75] [76] [77]  It is then possible to derive the evolution of W(x,p) with time and then deduce the transient evolution of the system. By differentiating W with respect to time.  By introducing the energy of the carriers in the bands we have **U = im\* v 2 + i k B T** + U, including the thermal energy.

$$U_q = -\frac{\hbar^2}{8m^*}\nabla^2 \ln(n)$$

In addition, the Uq is the potential Introduced by Iafrate, Grubin and Ferry [76] [77] . The Bohm potential can also be used, it is considered as a quantum potential in the former paper of David Bohm.

$$U_B = -\frac{\hbar^2}{2m\sqrt{n}}\frac{\partial^2 \sqrt{n}}{\partial x^2},$$

The quantum potential is expressed by the following formula:

$$U(\mathbf{x}) = \frac{-\hbar^2}{4m}\left[\frac{\nabla^2 P}{P} - \frac{1}{2}\frac{(\nabla P)^2}{P^2}\right] = \frac{-\hbar^2}{2m}\frac{\nabla^2 R}{R}. \qquad (8)$$

With this in consideration, the calculation of the transport equations is far less complicated. We have used this simulation scheme successfully in 2-dimentional simulations. The case of 3D is more complicated but cannot be easily simulated both because of the long simulations time if the case of SP and sometime failures of the simulation in the density gradient approach.

### 2.4.6.     Associated technologies

The way to fabricate these structures is now under thorough investigation. We have separated the two possible options. The most difficult question is the way to obtain a buried layer with the appropriate properties. The first option is ion implantation and the second is epitaxial growth. We can combine the two techniques, as we will show in a following paragraph.

### 2.4.6.1. Associated technologies : ion implantation

The most straightforward way to obtain the buried layer is to use ion implantation. The problem here is to get ions energies well above the 1MeV threshold in order to have a range in the material that does exceed the value of a few hundreds of nm. Implantation at lower energies induces a tail with the implanted impurity density at non-negligible values at the surface. High-energy ion implantation circumvent this effect at the expense of defect creation in the zone extending from the surface to the peak impurity concentration. We have made the following experiments.

   a)  1MeV Zn ion implantation for peak concentrations of $10^{18}$ cm$^{-3}$.
   b)  1 MeV Ge implantation for a peak concentration above $6\times10^{21}$ cm$^{-3}$
   c)  14 MeV P implantation for a peak concentration of $10^{14}$ cm$^{-3}$.

Table 8:

| Goal | Ion | Substrate | Energy | Doses | Substrate temperature |
|---|---|---|---|---|---|
| Implanted layer Q well | Ge Implantation | High resistivity silicon wafer 100 ohm.cm | 1MeV | $3 \times 10^{17}$ cm$^{-2}$ | Room temperature (300 K) |
| n-type buried layer | P implantation | High resistivity silicon wafer 100 ohm.cm | High energy 14 MeV | $5 \times 10^{12}$ cm$^{-2}$ | *Room temperature (300 K)* |
| Trapping layer Implanted layer | Zn implantation | High resistivity silicon wafer 100 ohm.cm 25 µm @ 5V SCZ 400pF/cm$^2$ | 1 MeV | $10^{14}$ cm$^{-2}$ | Room temperature (300 K) |
| Implanted layer Q well | Ge Implantation | Low resistivity silicon wafer 1 ohm cm 0.8 µm @ 5V SCZ 13 nF /cm$^2$ | 1MeV | $3 \times 10^{17}$ cm$^{-2}$ | Room temperature (300 K) |
| n-type buried layer | P Implantation | Low resistivity silicon wafer | High energy 14 MeV | $10^{13}$ cm$^{-2}$ | Room temperature (300 K) |
| Trapping layer Implanted layer | Zn Implantation | Low resistivity silicon wafer | 1 MeV | $10^{14}$ cm$^{-2}$ | Room temperature (300 K) |

Ion implantation with accelerators at these energies is not a standard material processing technique. Long processing times are necessary for the high doses required (Ge in Si).

Table 9: measured doses and fluxes

| Incident Angle 15 degrees | Measured Integrated Flux (electrodes) | Ion Measured | Flux (electrodes) cm$^{-2}$s$^{-1}$ | Duration |
|---|---|---|---|---|
| P | $5 \times 10^{12}$ cm$^{-2}$ | P$^{6+}$ | $1.17 \times 10^{10}$ | 7 mm 5 s |
| Ge | $2.28 \times 10^{17}$ cm$^{-2}$ | Ge$^+$ | $1.84 \times 10^{12}$ | 2070 mn |
| Zn | $1 \times 10^{14}$ cm$^{-2}$ | Zn$^+$ | $7.15 \times 10^{10}$ | 23 mn 20 s |

Table 10 : summarizing the SIMS results for the Zn implantations

| HR RBS Cyril Bachelet | SIMS BR Oxygen Incident Ions $^{64}$Zn | Background 1e16 cm-3 ?? Peak | SIMS HR Oxygen Incident Ions $^{64}$Zn | Background 1e16 cm-3 ??noisy Peak | SIMS HR Césium Incident ions | Background 2e17 cm-3 Peak |
|---|---|---|---|---|---|---|
| 0.0025 peak $1.25 \times 10^{20}$cm$^{-3}$ | A1 | $1.3 \times 10^{18}$cm$^{-3}$ I=$6 \times 10^{13}$cm$^{-2}$ | A1 (NI) | $10^{16}$cm$^{-3}$ I=$5.6 \times 10^{12}$cm$^{-2}$ | B1 | $2.1 \times 10^{18}$cm$^{-3}$ I=$1.14 \times 10^{14}$cm$^{-2}$ |
| | A2 | $1.2 \times 10^{18}$cm$^{-3}$ I=$6.2 \times 10^{13}$cm$^{-2}$ | A2 | $10^{18}$cm$^{-3}$ I=$5.22 \times 10^{13}$cm$^{-2}$ | B2 | $2.2 \times 10^{18}$cm$^{-3}$ I=$1.3 \times 10^{14}$cm$^{-2}$ |
| | A3 | $1.7 \times 10^{18}$cm$^{-3}$ I=$8.7 \times 10^{13}$cm$^{-2}$ | A3 | $1.2 \times 10^{18}$cm$^{-3}$ I=$5.7 \times 10^{13}$cm$^{-2}$ | | |
| | A4 | $4 \times 10^{17}$cm$^{-3}$ I=$2 \times 10^{13}$cm$^{-2}$ | A4 (up to peak) | $1.2 \times 10^{18}$cm$^{-3}$ | | |
| | A5 (NI) | $10^{16}$cm$^{-3}$ I=$2.2 \times 10^{12}$cm$^{-2}$ | | | | |
| HR ( peak) $10^{18}$cm$^{-3}$ | | | | | | |
| SRIM simulations | | Peak = $2.4 \times 10^{18}$cm$^{-3}$ | | | | Peak position 704 nm Measured: 722nm |

Table 11 : summarizing the SIMS results for the: Ge implantation

| $^{72}$Ge BR | Quantified Cs- Negative Secondary Ge et Si Ions  Peak position 647 nm Background: $10^{17}$cm$^{-3}$ | Quantified Cs+ ions MCs+ Positive Secondary Ions Peak position 551 nm Backgroundx $10^{17}$cm$^{-3}$ : | $^{72}$Ge HR | Quantified Cs+ Positive Secondary Ions Peak position 559 nm Background $1x10^{18}$cm$^{-3}$ : | $^{72}$Ge HR | Quantified Cs- initial negative Secondary Si Ge Ions Peak position at 658 nm Background : $1x10^{17}$cm$^{-3}$ | RBS : simulated measured spectrum (Cyril Bachelet) |
|---|---|---|---|---|---|---|---|
| C1 | $3x10^{21}$ cm$^{-3}$  I=1.25X10$^{17}$cm$^{-2}$ | | D1 | $1.35x10^{22}$cm$^{-3}$  I=5.0x10$^{17}$cm$^{-2}$ | C1 | ?? not available | HR : 7% $3.5x10^{21}$cm$^{-3}$ |
| C2 | $2.5x10^{21}$ cm$^{-3}$  I=1.08x10$^{17}$cm$^{-2}$ | | D2 | $1.35x10^{22}$cm$^{-3}$  I=5.04x10$^{17}$cm$^{-2}$ | C2 | ?? not available | BR : 9 % $4.5x10^{21}$cm$^{-3}$ |
| D1 | | $1.8x10^{22}$cm$^{-3}$  I=7x10$^{17}$cm$^{-2}$ | | | C3 | $2x10^{21}$cm$^{-3}$  I=8.5x10$^{16}$cm$^{-2}$ | |
| D2 | | $1.8x10^{22}$cm$^{-3}$  I=7x10$^{17}$cm$^{-2}$ | | | C4 | $6x10^{16}$cm$^{-3}$  I=2x10$^{12}$cm$^{-2}$ | |
| D3 | | No Signal | | | C5 | $2x10^{21}$cm$^{-3}$  I=8.64x10$^{16}$cm$^{-2}$ | |
| SRIM | Peak position: 646 nm | Peak = $4.58x10^{21}$cm$^{-3}$ | | | | | |

Table 12: summarizing the SIMS results for the phosphorous implantation

| A1 (test calib sample) | Peak (5e18cm-3) | Background : I=1x10$^{12}$cm$^{-2}$ | Integral $10^{14}$cm$^{-2}$ | Peak Position |
|---|---|---|---|---|
| A1W24004 (BR) | Peak(5x10$^{16}$cm$^{-3}$) | Background:I= 1x10$^{12}$cm$^{-2}$ | $3.4x10^{12}$cm$^{-2}$ | |
| A2W2404 (BR) | Peak(7x10$^{16}$cm$^{-3}$) | Background: I=1x10$^{12}$cm$^{-2}$ | $3.7x10^{12}$cm$^{-2}$ | |
| A3W2404 (BR) | Peak(7x10$^{16}$cm$^{-3}$) | Background:I= 1x10$^{12}$cm$^{-2}$ | $3.7x10^{12}$cm$^{-2}$ | |
| A4W2404 (BR) | Peak(No) | Background I=1x10$^{12}$cm$^{-2}$: | | |
| SRIM | Peak(4x10$^{16}$cm$^{-3}$) | | | 6.18 µm  6.00µm measured |

Table 13: Rapid summary of the results for Ge and Zn

| | Measured integrated Flux (SIMS) | Measured Integrated Flux (SIMS) | Measured Integrated Flux corrected from the 15 incident angle | |
|---|---|---|---|---|
| $^{64}$Zn | 0.545-0.57 x $10^{14}$ cm$^{-2}$ high resitivity 0.84 x $10^{14}$cm$^{-2}$ with Oxygen (low resitivity) | 1.14-1.3x$10^{14}$ cm$^{-2}$ high resitivity with Cs | 0.9998 x $10^{14}$ – 0.9657 x $10^{14}$ Zn cm$^{-2}$ | |
| $^{72}$Ge MCs+ | I=5.04x$10^{17}$cm$^{-2}$ | | 2.28x$10^{17}$ cm$^{-2}$ | |
| $^{72}$Ge-MCs- | | I=1.25X$10^{17}$cm$^{-2}$ | 2.28x$10^{17}$ cm$^{-2}$ | |

The simplest way to check the concentration profile is SIMS (Secondary Ion Mass Spectroscopy). The implantations were made at the EMIR/Jannus facility in Saclay. Most of the results have been published [67] [71]. The SIMS measurements give a profile that is close to the expected one. SRIM simulations. The quasi Gaussian distribution is observed with SIMS and the position of the peak concentration fits with the SRIM simulation. The difficult part concern the magnitude of the peak. The abrasive ion here is usually Cs (Caesium). The molecules are then then accelerated and analysed using mass spectrometry. For Phosphorous than is well characterized is silicon the correct value were found: For the Zn and Ge it was a bit more difficult. The concentration were corrected for the isotope proportions in natural element.

    The SRIM simulations were made corrected for a Td=20 eV and an incident angle of 15 degrees. The angle of incidence 15 degrees only accounts for a attenuation of a factor of 0.98 overestimation of the flux and integrated flux but has more important effect on the concentration profile. The Ge ($^{72}$Ge) concentration profile is not homogeneous with respect to position on the sample. The peak concentration measured was 25% of the atomic density (>$10^{22}$ cm$^{-3}$). This value was obtained after an alloy calibration method. The maximum value is below 50% of the atomic density, which is the peak value need for a correct operation of the projected device. The target-integrated flux was:

a) 3x$10^{17}$cm$^{-2}$ with an introduction rate of approximately 2x$10^{4}$cm$^{-1}$ this corresponds to a peak concentration is 6x$10^{21}$ cm$^{-3}$.
b) The SIMS measured dose was 2.29x$10^{17}$cm$^{-2}$ this corresponds to a peak concentration of 5.8x$10^{21}$ cm$^{-3}$.Or 12 % of the atomic density.
c) We have measured such value below this target with the Oxygen incident ions and above this target with the Cs so that the peak concentration lies between these two values. The difficulties for implanting at such high fluxes and duration main explain the discrepancies.
d) The positive MCs ions give higher value than the negative ones, MCs-

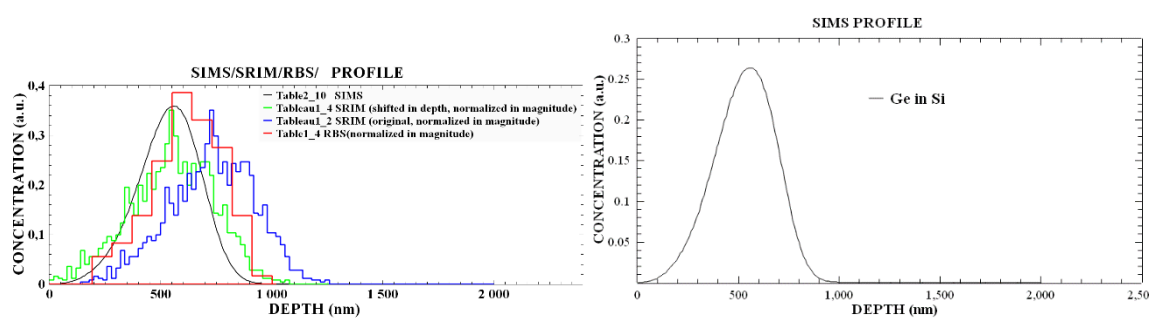

Fig, 40: Ge concentration profile (left compared with RBS and SRIM simulations), right SIMS only profile.

The expected Zn ($^{64}$Zn) integrated flux was to be $10^{14}$ cm$^{-2}$ corresponding to a peak concentration of $2.5 \times 10^{18}$ cm$^{-3}$, with an introduction rate of $2.5 \times 10^4$ cm$^{-1}$. The results of the SIMS measurements were:

a) The measured integrated flux was as expected $10^{14}$ cm$^{-2}$, using the electrode current. (see table )
b) There is a discrepancy between the integrated flus measured with the Cs, which higher than that measured with oxygen ion.
c) The peak concentration is on average: $1.83 \times 10^{18}$ cm$^{-3,}$ for Oxygen incident ions and $6.95 \times 10^{13}$ cm$^{-2}$ integrated flux, below the value both measured and targeted this is obtained on low resistivity samples. On the high-resistivity samples the peak concentration is: $1.2 \times 10^{18}$ cm$^{-3}$ corresponding to : $5.45 \times 10^{13}$ cm$^{-2}$
d) For Cs incident ions, the results on HR samples show an average integrated flux of: $1.22 \times 10^{14}$ cm$^{-2}$ and a peak concentration of: $2.15 \times 10^{18}$ cm$^{-3}$, above the measured values by 20 %. This could be explained by the presence in the secondary ions of species with the same Charge to Mass Ratio of the Ge secondary ions. Otherwise, although lower, the peak concentration corresponds to the target value by -11% to -24 % that is very good, if we consider that the implantation is inhomogeneous, which is probed by the electrode current.

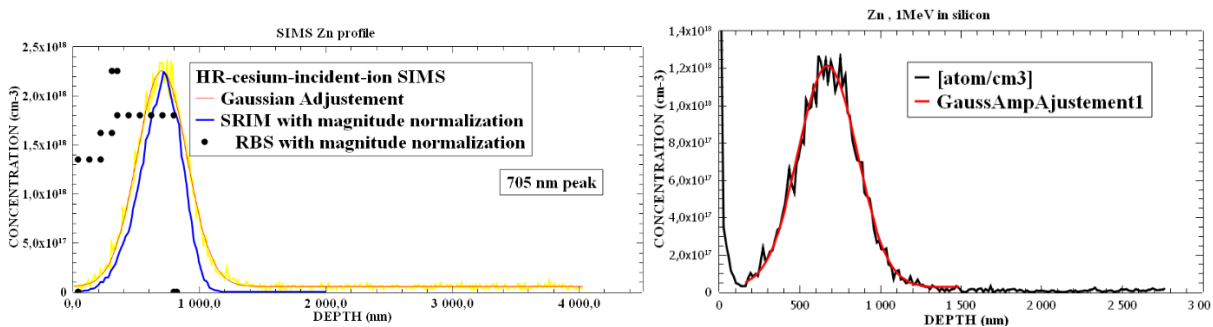

Fig. 41: Zn implantation profile (left resistivity 100 ohm.cm, right low resistivity 0.1 ohm.cm with Gaussian adjustment. On the left Rutherford Backscattering results and SRIM simulation.

The phosphorous ($^{31}$P) target integrated flux is $5 \times 10^{12}$ cm$^{-2}$ and the measured integrated flux with the electrodes : $5 \times 10^{12}$ cm$^{-2}$, (with an introduction rate of : $1.8 \times 10^4$ cm$^{-1}$ at 15 degrees. leads to a peak concentration of : $9 \times 10^{16}$ cm$^{-3}$, this is below the target : The results of the SIMS measurements are then :

a) The average measured integrated flux with Oxygen SIMS was: $3.91 \times 10^{12}$ cm$^{-2}$
b) The corresponding peak concentration was measured on average: $6 \times 10^{16}$ cm$^{-3}$ that is comparable to the target value obtained using the target implantation and the introduction rate derived from SRIM simulations. No measurements with Cs as the incident ions were made here.

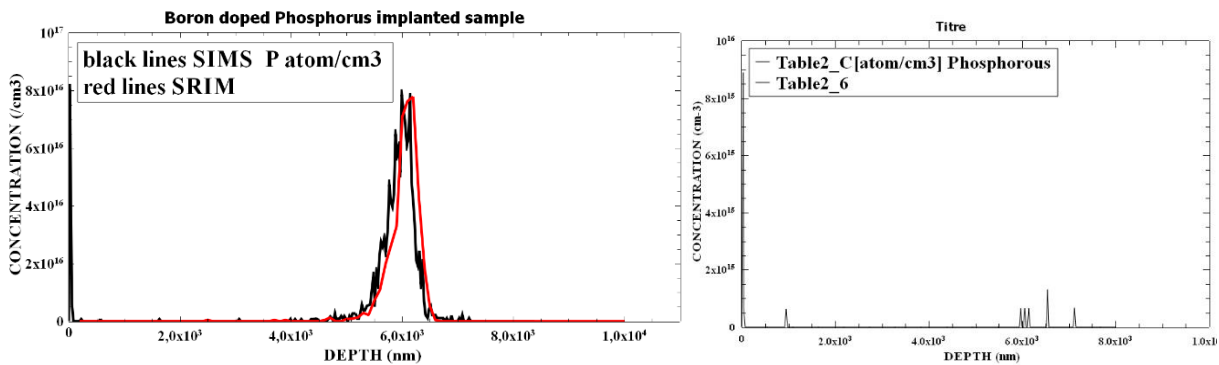

Fig.42: P distribution profile on a low resistivity samples (0.1 ohm.cm, (SRIM simulation and SIMS results). On the right non-implanted sample. Implantation energy: 14 MeV.

Fig.40 shows all the results of the high energy ion implantation, with the profile of the Ge concentration and the limitation of the technique

A conclusion that can be drawn is that the position of the peak concentration fits in all cases studied here with the SRIM simulations. This is true for the Zn, Ge and P ions. The density of implanted ions is controllable using these implantation techniques as well as the low energy standards ones. The problem comes from the defect control and the profile which is not abrupt enough be able to use such implanted structures such as the Ge and and Zn, However P implantation may be used at this energy as it.

## 2.4.6.2. Associated technologies : characterization

The main question is now related to the properties of these buried layers. The two possible problems are defects characterization and control within the implanted layer and above. The second is the stabilization of the layer to avoid any diffusion at room temperature that could prevent the stable operation of the device. For this purpose, we use the following experimental techniques.

a) Raman spectroscopy: this technique is non-destructive and give information on the coupling of the phonons with incident light.
b) Deep Level Transient Spectroscopy, which involves a MOS, PN, PIN, Schottky structure and give quantities related to the electrical properties of defects, capture cross-section, energy levels and densities.

**Raman spectroscopy results on the implanted layers:**

The defects and the impurities together with some other properties can be revealed by Raman spectroscopy if they have a high concentration or perturbate significantly the crystal lattice to add or substracted vibrational modes [78] [79] [80] [81] [82].

For silicon the range in the bulk silicon which contributes to the Raman signal is of the order of 500 nm above the peak concentration of the Ge and Zn ions (with the a laser wavelength of 488 nm in the green-blue region of the spectra) . For the Renishaw apparatus, the incident wavelength is 532 nm (green) according to the vendor.

First, the space group is Fd3m for silicon (diamond structure), the point group in Schoenefliess notation is $O_h$. Using the experimental results from numerous contributions and the making simulations from the Bilbao Crystallographic center, we have deduced that for the diamond structure, only one single first order mode is allowed. This corresponds in Mulliken notation to the following representations. The Mulliken symbol corresponds to the dimension of the character of the irreducible representation. T is of dimension 3 or triply degenerate. See Table 14-15 for silicon like structure.

$T_{2g}$, , which ( $\Gamma_5^-$) gives a triply degenerate mode at the center of the Brillouin zone. It is a first order optical mode. As there are 2 atoms per unit cell in silicon the number of phonon modes is 3x2 = 6 with 3 acoustic and 3 optical branches. The first order optical branches contribute to the first order Raman lines. When $T_{2g}$, modes are observed then the other modes are not $A_{2g,}$ and $E_{2g}$, modes depending on the incident light.

In this configuration the selection rules shows that the if we observe the $T_{2g}$, modes the other modes cannot be observed.

**Information of the Point Group O$_h$ (m-3m)**

**Character Table[1]**

| O$_h$(m-3m) | # | 1 | 4 | 2$_{100}$ | 3 | 2$_{110}$ | -1 | -4 | m$_{100}$ | -3 | m$_{110}$ | functions |
|---|---|---|---|---|---|---|---|---|---|---|---|---|
| Mult. | - | 1 | 6 | 3 | 8 | 6 | 1 | 6 | 3 | 8 | 6 | · |
| A$_{1g}$ | Γ$_1^+$ | 1 | 1 | 1 | 1 | 1 | 1 | 1 | 1 | 1 | 1 | x²+y²+z² |
| A$_{1u}$ | Γ$_1^-$ | 1 | 1 | 1 | 1 | 1 | -1 | -1 | -1 | -1 | -1 | · |
| A$_{2g}$ | Γ$_2^+$ | 1 | -1 | 1 | 1 | -1 | 1 | -1 | 1 | 1 | -1 | · |
| A$_{2u}$ | Γ$_2^-$ | 1 | -1 | 1 | 1 | -1 | -1 | 1 | -1 | -1 | 1 | · |
| E$_g$ | Γ$_3^+$ | 2 | 0 | 2 | -1 | 0 | 2 | 0 | 2 | -1 | 0 | (2z²-x²-y²,x²-y²) |
| E$_u$ | Γ$_3^-$ | 2 | 0 | 2 | -1 | 0 | -2 | 0 | -2 | 1 | 0 | · |
| T$_{2u}$ | Γ$_5^-$ | 3 | -1 | -1 | 0 | 1 | -3 | 1 | 1 | 0 | -1 | · |
| T$_{2g}$ | Γ$_5^+$ | 3 | -1 | -1 | 0 | 1 | 3 | -1 | -1 | 0 | 1 | (xy,xz,yz) |
| T$_{1u}$ | Γ$_4^-$ | 3 | 1 | -1 | 0 | -1 | -3 | -1 | 1 | 0 | 1 | (x,y,z) |
| T$_{1g}$ | Γ$_4^+$ | 3 | 1 | -1 | 0 | -1 | 3 | 1 | -1 | 0 | -1 | (J$_x$,J$_y$,J$_z$) |

**Raman Tensors**

| | A$_{1g}$ | | | E$_g$ | | | E$_g$ | | | T$_{2g}$ | | | T$_{2g}$ | | | T$_{2g}$ | |
|---|---|---|---|---|---|---|---|---|---|---|---|---|---|---|---|---|---|
| a | · | · | b | · | · | -3$^{1/2}$b | · | · | · | · | · | · | · | d | · | d | · |
| · | a | · | · | b | · | · | 3$^{1/2}$b | · | · | · | d | · | · | · | d | · | · |
| · | · | a | · | · | -2b | · | · | · | d | · | d | · | · | · | · | · | · |

Table 14: Table of characters of the O$_h$ group (left) and table of the allowed first order Raman modes (right)

**Raman Active Modes**

| WP | A$_{1g}$ | A$_{1u}$ | A$_{2g}$ | A$_{2u}$ | E$_u$ | E$_g$ | T$_{2u}$ | T$_{2g}$ | T$_{1u}$ | T$_{1g}$ |
|---|---|---|---|---|---|---|---|---|---|---|
| 16c | · | · | · | · | · | · | · | · | · | · |
| 8a | · | · | · | · | · | · | · | 1 | · | · |
| 8b | · | · | · | · | · | · | · | 1 | · | · |

**Second Order Raman Activity**

In the next table one can find the information the second order Raman active modes:

| ⊗ | A$_{1g}$ | A$_{1u}$ | A$_{2g}$ | A$_{2u}$ | E$_u$ | E$_g$ | T$_{2u}$ | T$_{2g}$ | T$_{1u}$ | T$_{1g}$ |
|---|---|---|---|---|---|---|---|---|---|---|
| A$_{1g}$ | A$_{1g}$ | A$_{1u}$ | A$_{2g}$ | A$_{2u}$ | E$_u$ | E$_g$ | T$_{2u}$ | T$_{2g}$ | T$_{1u}$ | T$_{1g}$ |
| A$_{1u}$ | · | A$_{1g}$ | A$_{2u}$ | A$_{2g}$ | E$_g$ | E$_u$ | T$_{2g}$ | T$_{2u}$ | T$_{1g}$ | T$_{1u}$ |
| A$_{2g}$ | · | · | A$_{1g}$ | A$_{1u}$ | E$_u$ | E$_g$ | T$_{1u}$ | T$_{1g}$ | T$_{2u}$ | T$_{2g}$ |
| A$_{2u}$ | · | · | · | A$_{1g}$ | E$_g$ | E$_u$ | T$_{1g}$ | T$_{1u}$ | T$_{2g}$ | T$_{2u}$ |
| E$_u$ | · | · | · | · | A$_{1g}$+A$_{2g}$+E$_g$ | A$_{1u}$+A$_{2u}$+E$_u$ | T$_{2g}$+T$_{1g}$ | T$_{2u}$+T$_{1u}$ | T$_{2g}$+T$_{1g}$ | T$_{2u}$+T$_{1u}$ |
| E$_g$ | · | · | · | · | · | A$_{1g}$+A$_{2g}$+E$_g$ | T$_{2u}$+T$_{1u}$ | T$_{2g}$+T$_{1g}$ | T$_{2u}$+T$_{1u}$ | T$_{2g}$+T$_{1g}$ |
| T$_{2u}$ | · | · | · | · | · | · | A$_{1g}$+E$_g$+T$_{2g}$+T$_{1g}$ | A$_{1u}$+E$_u$+T$_{2u}$+T$_{1u}$ | A$_{2g}$+E$_g$+T$_{2g}$+T$_{1g}$ | A$_{2u}$+E$_u$+T$_{2u}$+T$_{1u}$ |
| T$_{2g}$ | · | · | · | · | · | · | · | A$_{1g}$+E$_g$+T$_{2g}$+T$_{1g}$ | A$_{2u}$+E$_u$+T$_{2u}$+T$_{1u}$ | A$_{2g}$+E$_g$+T$_{2g}$+T$_{1g}$ |
| T$_{1u}$ | · | · | · | · | · | · | · | · | A$_{1g}$+E$_g$+T$_{2g}$+T$_{1g}$ | A$_{1u}$+E$_u$+T$_{2u}$+T$_{1u}$ |
| T$_{1g}$ | · | · | · | · | · | · | · | · | · | A$_{1g}$+E$_g$+T$_{2g}$+T$_{1g}$ |

Active   Inactive   **bold** Raman active modes

Table 15: first order selection rules (left) and (right) second order selection rules allowed modes in bold

There are one first order Raman mode and three-second order modes. One of the second order mode is an overtone of the first order 3-degenerate mode. This main Raman mode in Silicon at first order is the 520 cm$^{-1}$ line in Stokes mode which is due to the triply degenerate optical phonon O(Γ) center of Brillouin zone k=0 [83]. Second order modes allowed for the silicon give rise to the lines observed in the spectrum. In Germanium due to the differences in the elastic constant (Hooke Law) and lattice constants, this mode is observed at lower frequency shifts at 300 cm$^{-1}$.

### $^{31}$P (only) implanted sample (100):

We find the same lines as in the as-grown silicon:

They are listed as so : 70 cm$^{-1}$ (difficult to resolve, due to the filter), 300 cm$^{-1}$ and the 520 cm$^{-1}$ which characteristic of a Si-Si mode, a 620-670 cm$^{-1}$ which is a weak feature, the 940 cm$^{-1}$ and 970 cm$^{-1}$ double feature also observed in as-grown material at 430 cm$^{-1}$ 640 -970 cm$^{-1}$, slight shift to lower energies. The line at 300 cm$^{-1}$ is a 2TA mode of higher order found in as-grown silicon. Other features such as the 250 cm$^{-1}$ mode is difficult to resolve, as it is a weak signal as well as in as-grown samples. The Raman lines below the 520 cm$^{-1}$ line are commonly ascribed to second order modes and are not well identified. In [84]]. Transverse Acoustical modes are found in the implanted sample (they should be 2$^{nd}$ order or higher).

The residual silicon dioxide present in the upper part of sample should not contribute to the spectrum as it has a few nanometers in thickness. The silicon dioxide is transparent for visible light. Comparison of the P implanted sample with the reference sample or the spectrum obtained on the non-implanted zone of the sample show no shift for the characteristic: 520 cm$^{-1}$ line, which is characteristic of crystalline silicon. With the present resolution, this means that in the sample depth investigated no compressive or tensile strain can be observed in the silicon matrix. The conclusion is that the P implanted zone is out of reach of the Raman investigation, being located 6 µm or more below the surface. The defects created by the implantation have no influence on the spectra because of the dose is too low. SRIM simulations support this conclusion.

The average introduction rate determined by the SRIM simulations is of the order of 0.1 displacements per angstrom x ion : for a target depth of 6 microns, this is 6000 nm or 60000 Angstroms this means that each ion displaces 6000 atoms in the zone above the peak P concentration (9174 in total). This leads to 6000 atoms / surface unit x integrated flux, or written differently 6000 atoms /volume unit x integrated flux = introduction rate. For an integrated flux of $5 \times 10^{12}$ cm$^{-2}$ the concentration of primary displacements is of the order of : 30000 x $10^{12}$ cm$^{-2}$, hence: $3 \times 10^{16}$ cm$^{-3}$ in the region above the phosphorous peak concentration. The consequence of this is that in the region down to 500 nm below the surface the introduction rate for the vacancies is of the order of 1000 cm$^{-1}$ that yields a defect concentration below the $10^{16}$ cm$^{-3}$ limit. In this case, DLTS should be used to characterize the defects, as for the low resistivity samples the doping level is significantly larger at around $10^{17}$ cm$^{-3}$. This will be made in a further work.

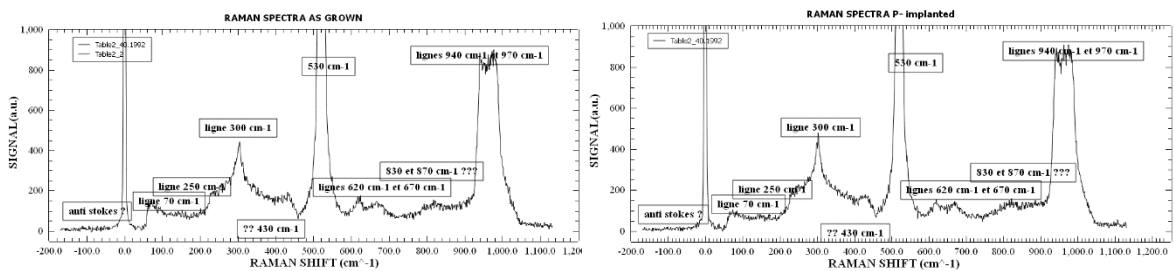

Fig.43: Raman spectra in Stokes mode of the as-grown silicon sample (left) and on the P implanted sample (right)

**$^{64}$Zn (49 %, in natural abundance, (ZnO output of the cathode) implanted sample (100):**

In 1-MeV Zn implanted samples the peak concentration is close to $10^{18}$ cm$^{-3}$ and the peak is located ~600 nm below the surface making the contribution of the Zn and the implantation induced defects significant. This has major consequences. First the 960-970 cm$^{-1}$ overtone doublet is much reduced in magnitude and the twofold Lorentzian is now less clear. The 520 cm$^{-1}$ line sees its magnitude much reduced, this perhaps originating from a high defect density in the Si layer investigated. The spectra show a feature at 300 cm$^{-1}$ with a similar amplitude with similar lines at 100 cm$^{-1}$ and are the same as the ones found in as-grown silicon with a higher magnitude. A shoulder appears on the left of the 520 cm$^{-1}$ line which origin is unknown. Although this line seems to be slightly shifted to the left (this could be the effect of this shoulder) the effect is too small to be accounted for, given the present resolution of the spectrum. The 100 cm$^{-1}$ may be attributed to a Zn (Si-Zn) mode as the $^{64}$Zn (substitutional) has a higher mass than the $^{28}$Si atom the frequency should be lower that the Si-Si mode (I shall discuss this conclusion in a future work ). The 960-970 cm$^{-1}$ overtone line has a very low magnitude (divided by 75), and the same can be said of the magnitude of the characteristics line 520 cm$^{-1}$ line,) factor of 2000).

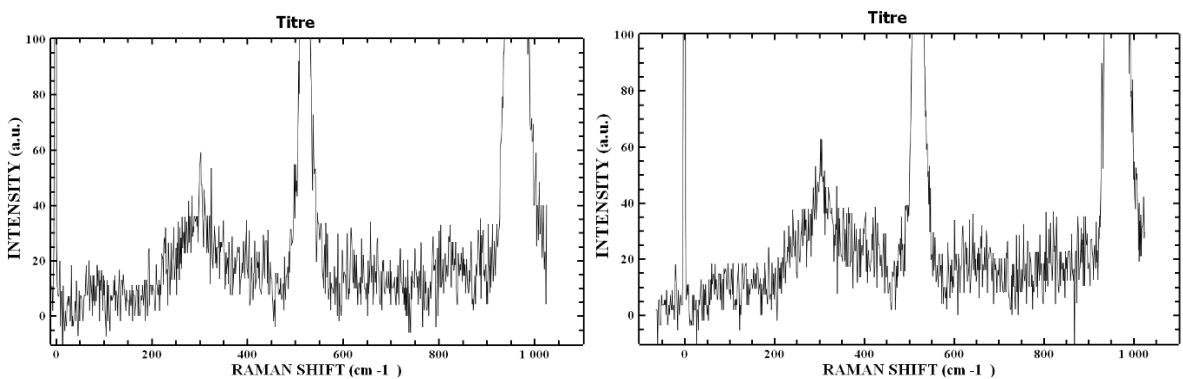

Fig.44: Raman lines of Zn implanted Silicon

**Natural Ge ($^{72}$Ge, $^{74}$Ge) implanted sample (100):**

**Germanium Raman Lines:**

The Raman spectrum of the natural germanium single crystals have been publish as early as 1967 [85]. As in silicon, the Raman spectrum at first order reduces to a single line at 300 cm$^{-1}$ this corresponds to the 520 cm$^{-1}$ line in silicon. This is always at room temperature [86].

Weinstein and Cardona have identified the second order germanium lines in 1973 [87]. There is a shoulder triple line feature 550-660 cm$^{-1}$ which is an overtone of the 300 cm$^{-1}$ first order line feature (2 phonons)

There are also a triple line shoulder in the 100-150-200 cm$^{-1}$, which was observed and is due to a second order mode as it was seen in silicon.

Now we may study the Ge implanted silicon as being first a material close to a SiGe alloy with because of the near surface initial disorder due to the implantation has a very high defect concentration. Some constraints due to lattice mismatch should exist.

**SiGe alloys:**

Raman spectra on SiGe strained layer epi structures have been published since 1989 [86] [88] [78] [89]. Prior to this Feldman Ashkin and Parker found a dependence of the 1O mode in SiGe alloys with this formula: As the SiGe is a disordered alloy it can be described by the usual way. A Modified Random Element Isodisplacement Model is used in these cases. Chang and Mitra [90] introduced it in 1968-1971 for mixed crystals. This is simple in fact as it takes into account the proportion x of the specie A of a mixed crystal : $A_xB_{1-x}$ where B is the other species. The force constants are not determined by a linear relation with x but by a condition at the limit x=0 and x=1. This model applies to the long wavelength optical modes. K close to zero.

Raman shift =389+/- 2 + (0.5+/-0.1) x cm$^{-1}$ where x is the silicon concentration in percentage. This is not true in more recent studies in which the Ge-Ge optical mode is found in SiGe alloys in a wide concentration range. Alonzo and Winer found the 280 cm$^{-1}$ line in SiGe that remains detectable in material having Ge proportions higher than 28 % . This line in not present in our as-implanted samples. More recently O. Pages, J. Souhabi,. J. B. Torres,.A V. Postnikov and K. C. Rustagi [88] studied the SiGe alloys with Raman Spectroscopy.

For these alloys according to Pages and al. the mode shifts linearly from 285 cm$^{-1}$ to 300 cm$^{-1}$ when the Ge proportion x shifts from 20 % to 100 % of the atomic density.

The Si-Si mode follows a linear shift from the 520 cm$^{-1}$ of the line to 460 cm$^{-1}$ when the Ge proportions vary from 0 to 100%.

The Si-Ge mode has its line located at: 406 cm$^{-1}$ for 50 % Ge in the silicon matrix. For x=10% line is located at 400 cm$^{-1}$ and x=95% the line is located at 395 cm$^{-1}$ .

With this in mind, we could determine the Ge concentration in the probed layer.

**Implanted Samples:**

The presence of a line at 450-460 cm$^{-1}$ should be interpreted by a the presence of a Si-Si mode in a surrounding lattice with a concentration of Ge close to 100% .However, this line is dominant in this sample, which should not be the case in a region with high Ge concentrations.

However, the presence of a line at 350 cm$^{-1}$ could suggest that the second order mode of Si are still visible. These are second order acoustical modes. This is difficult to understand as the overtone mode at 970 cm$^{-1}$ has vanished. The acoustical modes should be less sensitive to the disorder.

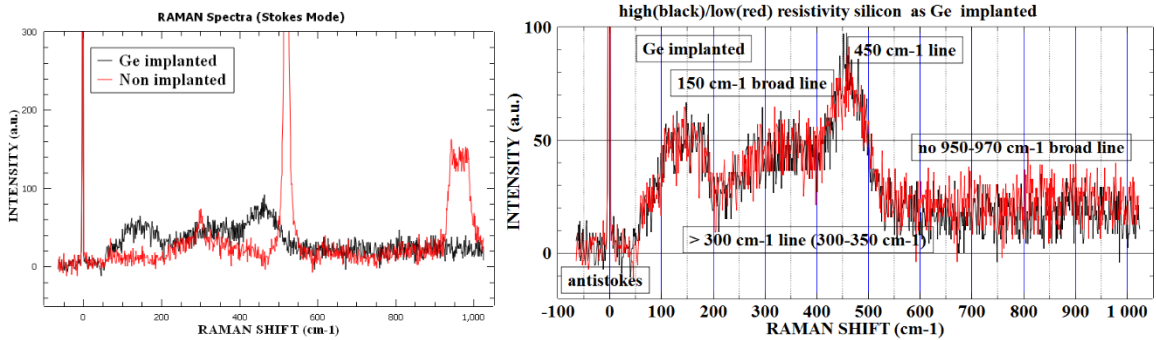

Fig.45: Raman spectrum of the Ge implanted silicon.

We have performed a thermal anneal at 850 degrees Celsius to remove the near surface defects

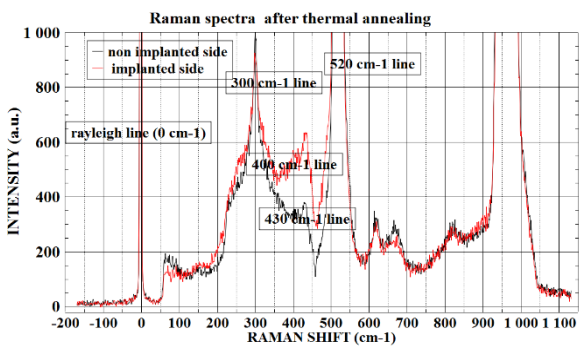

Fig.46: Raman spectrum of the Ge implanted and annealed silicon.

The only marked difference in the spectra is the existence of a 430 cm$^{-1}$ line in the implanted and annealed sample. This line could be due to the Si-Ge mode with 50 % Ge proportion but some explanation should be given for the shift to higher frequencies. Therefore the effect of both disorder and strained should be introduced in the analysis.

**Quantitative analysis and interpretations:**

The paper from Renucci et al. [91] [92] has enabled the determination of the Grüneisen coefficients in the silicon-germanium alloys. The results are as follow in $Si_{0.16}Ge_{0.84}$. So:

a) Ge-Si mode at 401 cm$^{-1}$ : $\gamma=1.2$ +/-0.1
b) Ge-Ge mode at 297 cm$^{-1}$ : $\gamma=1.13$ +/0.1
c) Si-Si mode at 520 cm$^{-1}$ : $\gamma=1.13$ +/-0.11 in silicon.

When the pressure is increased then the Raman shift is increased so it is straightforward to conclude that the Raman shift increases when the lattice parameter reduces (hydrostatic pressure used so that the effect is isotropic). In addition, SRIM simulations show that the material should be heavily damaged in the first 1 µm below the surface. It has been shown that the first 500 nm affects the Raman spectra. It was shown too that a thermal treatment up to 800 degrees anneals all the neutron-induced defects. We have to take into consideration two effects. The disorder due to displacements and the introduction of Ge atoms have to be considered with the effect of strain in the implanted layer.

**450 cm$^{-1}$ line: observed in the as-implanted sample.**

Some tailing towards the left (low shifts) is observed for the 450-500 cm$^{-1}$ but the may effect is to make the 520 cm$^{-1}$ first order line vanish. A 450-500 cm$^{-1}$ single Lorentzian shape, this shape is close to the silicon line.

It was shown that the shapes found in SiGe with a low Ge proportion (weak peaks), are the 50-100 cm$^{-1}$ (difficult to distinguish signal) as well as a 350 cm$^{-1}$ signal. Other authors have observed a 95 cm$^{-1}$ and 490

cm$^{-1}$ line feature in neutron irradiated silicon, the first being close to the very weak 50-100 cm$^{-1}$ feature found in our samples, and of course the 450-500 cm$^{-1}$ single Lorentzian shape.

We have also observed the presence of a broad 150 cm$^{-1}$ (2 lines) feature.

**430 cm$^{-1}$ line: observed in the annealed sample.**

The 430-400 cm$^{-1}$ line it can be ascribed to Si-Ge mode.

As the Grüneisen coefficients are given by: $\frac{\partial \omega}{\partial V}\frac{V}{\omega} = -\gamma$ and that $\gamma = 1.2$ we can deduce that 430 -400 =30 cm$^{-1}$ and then $\frac{\Delta \omega}{\omega} = 0.075$ and $\frac{\Delta V}{V} = \frac{0.075}{-1.2} = -0.063$ At first order this gives the following relation: $3\frac{\Delta r}{r} = -0.063$ and hence: $\frac{\Delta r}{r} = -0.021 = -2.1\%$

This should mean that the zone that contributes to the Raman signal is compressively strained. The 300 cm$^{-1}$ line is near identical to the lines found in un-implanted silicon. The line (TA high order) is unchanged by implantation as it was found in neutron-irradiated silicon. The 450-500 cm$^{-1}$ single Lorentzian shape appear in the spectra as we have observed in our samples.

In Ge implanted Si the 470 cm$^{-1}$ can be attributed to the same mode as in amorphous silicon. We may conclude that the 450-500 cm$^{-1}$ shape observed in our un-annealed samples is due to two the superposition of:
   a) The shift of the 520 cm$^{-1}$ Si-Si line due to the strain in the upper part of the region (tensile in this case, shift to lower wavevectors) and the disorder.
   b) The Ge-Si line which is due to the buried layer with high germanium concentration which is compressively strained

As we have shown the 330 cm$^{-1}$ large line can be attributed to Ge-Ge mode with the effect of disorder the shift to higher wavenumber being the result of the compressively strain GeSi layer in the GeSi layer buried layer. The Grüneisen coefficient for SiGe is of the order of g= -1.2 for this SiGe material and does not strongly depend on the composition:
$$\frac{\partial \omega}{\partial V}\frac{V}{\omega} = -\gamma$$

The shift is 330 -300 =30 cm-1, thus $\frac{\Delta \omega}{\omega} = -0.09$ and $\frac{\Delta V}{V} = -0.09$ and if we consider the volume of a sphere or any volume shape, the radius r (or linear size) at first order varies as:
$$3\frac{\Delta r}{r} = -0.09$$

This leads to: $\frac{\Delta r}{r}$= -0.03

**Conclusions:**

This small demonstration shows that the lattice should be compressively strained in the volume with a Ge proportion is at its highest of a few tens of percents. The lattice constant is reduced by up to 2 %.

The lattice mismatch between Si and Ge is be: (5.64613(Ge)-5.43095(Si))/5.43095 = 0.0396763=4 % of the silicon lattice constant.

This means that Ge should be compressively strained, and Si tensely strained. Our results show that the SiGe is compressively strained which is favourable for the operation of the Quantum Well. because in this case the formation of a Valence Band Deep Quantum well of 0.30-0.35 eV depth is possible.

**DLTS results on the implanted layers:**

For the high-energy Zn ion implantation some DLTS results have been obtained on Schottky contacts formed by the evaporation of a Al layer on the p-type high resistivity samples. A bad rectifying structure is obtained but a capacitance versus voltage plot can be measured, and with some DLTS spectra were obtained (credit F. Olivie LAAS). Prior to this, DLTS measurements were made on my home made DLTS system in

Saclay on HPGe samples. It is planned to use this system to characterize the defects on the P implanted samples.

The results show the existence of substitutional Zn with the characteristic deep acceptor level. The other shallower level is screened/quenched and cannot be observed in our spectra. The Zn implanted layer has a defect concentration, which is close to the net doping level. As no thermal anneal was made on the samples the concentration profile and the local defect configuration may be sensitive to applied electric field. This means that no quantitative conclusion may be made at this stage of the study. The strain on the upper part of the sample can also be established. These technologies could be usable for large structures with a buried gate of around 100 nm in thickness and one-micron device size. This is about one order of magnitude above the goal of a 100nm pitch pixel.

### 2.4.7. Future: Epitaxial growth and outlook

The use of ion implantation is not the sole option for buried layer fabrication. The other way to obtain a thin ( =< 20 nm) SiGe layer is to grow an epitaxial layer of Ge on a buffer layer of Si on the silicon substrate. This is now under study with partners at C2N [93] [94].

This technique will be more adequate than ion implantation for the future structures with thin buried gates. Moreover, the defect control is easier to make than in implanted layers. In addition to this, the condensation technique is under evaluation in our implanted Ge layers. It consist in the use of a thermal oxidation of a SiGe layer. As this thermal oxidization acts preferably on the Silicon atoms, the near oxide/interface becomes depleted in Si and enriched in Ge. This technique is used in industrial applications. We are now investigating its use for the realization of Ge enriched and Si epitaxially grown SiGe layer.

The use of CVD to epitaxially grow Ge on Si and Si on Ge structures is now fully under investigation, with the participation of the C2N and other CNRS laboratories.

## 3. Conclusions:

The results we have attained in the development of the SiGe buried gate could allow the fabrication of a large TRAMOS/DOTPIX structure such as this one with a 1-10 micrometer pitch and with a thickness less than 100 micrometers The buried gate should then be located then be approximately 6-10 micrometers below the surface. Such a structure would need simulation work to check its final characteristics. However, because of its future dimensions, we are now more interested in the development of epitaxially growth structures with the performance as we have already simulated. No definitive conclusions may be drawn straight away. The development of epi-structures is now under study and will continue in the near future.

Note 1: In a TPC detector, the aim is the same the reconstruction of the track is possible trough the time dependence of the current signal, which is collected at spatially distributed electrodes

Note 2: The pseudorapidity η is defined by : η=–ln(tan(θ/2))

Note 3 : MIMOSA stands for (Minimum Ionizing particle MOS Active pixel sensor)

Note 4 / MAPS stands for Monolithic Active Pixel Sensor

Note 5: it would be preferable to study the annealing process at temperatures close to the operation temperature, especially if some reverse annealing may occur. This is a valid conclusion for the future designs.

Note 6: using GaAs would be interesting to investigate but this is a specific field of research

# Acknowledgments:

.

The initial contribution of the Strasbourg group is acknowledged here. Past contribution with some importance came from Gregorz Deptuch (IPHC),Wojciech Dulinski (IPHC), Claude Colledani (IPHC), Abd-el-Kader Himmi (IPHC), Auguste Besson (IPHC), Marc Winter (IPHC) as the coordinator and Christine Hu (IPHC) , Fabrice Guilloux (IPHC then) and Gilles Claus(IPHC), Michael Deveaux and Michal Szelezniak (IPHC then).

In IRFU/Saclay contributed in these work co-workers such as Yavuz Degerli (first as a post-doc and permanent staff member), Marc Besancon and Pierre Lutz. Our PhD student Yan Li contributed to the development of the MIMOSA 8 chips for instance through test codes.

The EMIR/ Jannus facility (Saclay) contributed to the implantation study. I acknowledge the contributions of Yves Serruys and Frederic Lepretre. The help Gaelle Gutierrez for Raman measurements was of strong importance. François Jomard of the University of Versailles-Saint Quentin contributed with his SIMS measurements.

In LAAS Toulouse, the implication of Pr. Francois Olivié has allowed some DLTS measurements to be done. Eric Imbernon has allowed some annealing and oxidization to be made on implanted Si samples

For future work I must count on Charles Renard, Geraldine Hallais (C2N)